# Computing planetary atmospheres with algorithms derived from action thermodynamics and a novel version of the virial theorem for gravitating polyatomic molecules

Ivan R. Kennedy (Faculty of Agriculture and Environment, University of Sydney, Australia)

**\*Correspondence to:** ivan.kennedy@sydney.edu.au

An objective revision of Laplace's barometric formula for isothermal planetary atmospheres is proposed. From Clausius' virial theorem equating the root-mean-square kinetic energy to half the gravitational potential energy, planetary atmospheres are required to have declining temperature with altitude as a consequence of the interaction between thermodynamic heat flow and gravity. The virial-action hypothesis predicts non-adiabatic lapse rates in temperature yielding a practical means to calculate variations with altitude in atmospheric entropy, free energy, molecular density and pressure. Remarkably, the new formulae derived enable prediction of atmospheric profiles with physical properties closely resembling those observed on Earth, Venus and Mars. These new formulae provide an objective basis for computing the dynamic morphology of the atmosphere. Climate scientists may consider this explanatory hypothesis for self-organisation of planetary atmospheres for its possible relevance for predicting global surface temperatures.

**Introduction**

The main hypothesis advanced in this paper is that algorithms developed from the physical property of action with Clausius' virial theorem can enable plots of the atmosphere's dynamic morphology. This hypothesis is based on the logarithmic relationship between action (Kennedy, 2000, 2001; Rose *et al*., 2008) and entropy (Kennedy et al., 2015) and a steady-state equilibrium in the atmosphere between heat flow to space and gravity. However, as with any hypothesis, this objective model needs to be rigorously tested against available evidence and critical experimental tests proposed (Popper, 1972).

The Laplace barometric formula for the atmosphere is typically given as:

$$p_h/p_o = n_h/n_o = e^{-mgh/kT} \qquad (1)$$

Here in equation (1), the gas pressure at an altitude $h$ ($p_h$) is compared to that at the base level ($p_o$). This is frequently given as a ratio of the number of molecules per unit volume at each altitude $h$ ($n_h/n_o$).

So, taking natural logarithms, we can write

$$kT\ln(n_h/n_o) = -mgh \qquad (2)$$

This gravitational formula was confirmed in 1910 for isothermal conditions using Brownian particles by Perrin (1909), who also verified Einstein's theory of translational and rotational Brownian motion. For this work, the Parisian chemist was awarded the Nobel prize for his reassuring experimental demonstration of molecular reality close to the year Einstein received his prize for his explanation of the photoelectric effect of light in ejecting electrons from a metallic surface. This classical work of Einstein and Perrin on Brownian motion provided the original inspiration for developing action thermodynamics (Kennedy, 2000, 2001).

In the action theory, this formula was considered as indicating an entropy or free energy change accompanying the change in the action volume for each molecule between zero altitude and the height $h$, as shown in equation (2). Alternatively, the transfer of a molecule from zero altitude to $h$ constitutes a change in gravitational potential equivalent to $mgh$. However, despite a considerable mathematical tradition regarding its derivation (Berberan-Santos, 1997), the well-known flaw of the Laplace formula is that it fails to correspond to reality regarding absolute temperature. The formula ignores the marked decline in temperature with altitude in the troposphere – an observation that any student of the flight path screen on modern airliners is well aware.

In this paper, the temperature gradient with altitude is proposed to be primarily linked to the dynamic interaction between atmospheric thermodynamics and gravitational work. Furthermore, this temperature gradient would necessarily exist in a static atmosphere even in the absence of adiabatic expansion or compression. We have shown (Kennedy et al., 2015) that the total thermal energy required by a molecular system termed here the entropic energy can be expressed as the product (*ST*) of its entropy (*S*) and its absolute temperature (*T*). As defined by Clausius (1875), this can be composed of sensible heat directly affecting temperature as well as work-heat, which does not. Thus, for a mole of monatomic molecules in a canonical ensemble, as defined by Gibbs (1902), the following concise relationship involving a function of relative action (@/$\hbar$) was shown to hold. The action term involved (@) for translation, equal to $(3kTI_t)^{1/2}$, where $I_t$ indicates the translational inertia ($I_t = mr_t^2$), explained in more detail following.

$$ST = RT\ln[e^{5/2}(@/\hbar)^3/z_t] \qquad (3)$$

The total field thermal energy, kinetic and potential, indicated by the entropy $R\ln[e^{5/2}(@/\hbar)^3/z_t]$ at the temperature $T$ is simply their product $RT\ln[e^{5/2}(@/\hbar)^3/z_t]$; $z_t$ is a correction factor for symmetry. As a result, for monatomic molecules argon, neon or helium in the atmosphere, based on temperature change alone, we might expect to have the following difference in absolute thermal energy content *sT* per molecule at the Earth's surface and at an altitude $h_n$.

$$s_o T_o - s_n T_n = kT_o\ln[e^{5/2}(@_{to}/\hbar)^3/z_t] - kT_n\ln[e^{5/2}(@_{tn}/\hbar)^3/z_t] \qquad (4)$$

This paper will show that this equation needs to be mathematically adjusted to account for the conversion of thermodynamic heat to gravitational energy. It achieves this with the aid of a second novel application – that of the virial theorem to the atmosphere as a gravitationally bound system of particles. These two novelties lead to highly precise barometric equations for an atmosphere in steady state equilibrium with thermal energy flow.

**Methodology and Results**

Action thermodynamics and its utility for calculating entropy and free energy has been explained in our companion publication (Kennedy et al., 2015). However, the virial theorem has not been considered until now as relevant to the Earth's atmosphere. Here we will argue that, just as in predicting the evolution of stars, the virial theorem can be usefully applied, by assuming that morphogenesis of planetary atmospheres can be regarded as thermodynamically reversible.

*The Virial Theorem*

Justification for the equations relating thermodynamics to gravity that follow is provided by the virial theorem as given by Clausius in 1870. According to this theorem, for reversible or steady state processes in a central force field such as a gravitational field, the time-averaged magnitude of the root-mean square kinetic energy is equal to half the time-averaged potential energy. Clausius stated that "the mean *vis viva* of the system is equal to its virial", implying that the time-averaged kinetic energy is equal to half the average potential energy (note that the *vis viva* term as used in the 17th and 18th centuries usually referred to twice the kinetic energy). The same conclusion is sometimes derived from Lagrange's identity of the 18th century as a time-average of the second derivative of the moment of inertia ($I = \Sigma m r_i^2$) with respect to time (equivalent to the first derivative of the action) being zero. Thus, for *T* as kinetic energy and *V* as potential energy, we have

$$\tfrac{1}{2} d^2 I / dt^2 = 2T + V = 0$$

and so

$$2T = -V$$

The theorem has been applied to the evolution of stars and is also illustrated aptly for our purpose in Bohr's model of the hydrogen atom, where the reversible absorption or emission of a quantum of energy is matched by reversible decreases or increases respectively in the electron's kinetic energy of exactly the same magnitude. The change in potential energy of the electron is thus twice this magnitude and opposite in sign to the change in kinetic energy, on a statistical basis.

In his lectures on gas theory, Boltzmann (1896) presented a succinct description of Clausius' concept of the virial for gases as derived as follows.

"Let $m_h$ be the mass of a point and $x_h, y_h, z_h, c_h, u_h, v_h, w_h$ be its rectangular coordinates, its velocity and its velocity components along the coordinate axes respectively, at some time point *t*. Let $\xi_h, \eta_h, \zeta_h$ be the components of the total force acting on this material point at the same time...……". Boltzmann credits Clausius as establishing that "No matter how long a time of motion may be chosen, the absolute value of any coordinate or velocity component must remain smaller than a fixed finite quantity, which has the value *E* for the coordinate and the value ε for the velocity components. Such molecular motions giving rise to thermal phenomena we will call stationary". The stationary property thus belongs to the physical dimension of momentum by position, or action. "By virtue of the equations of motion of mechanics, we have:

$$m_h(du_h/dt) = \xi_h$$

Hence

$$d(m_h x_h u_h)/dt = m_h u_h^2 + x_h \zeta_h$$

If one multiplies this equation by $dt$, integrates over an arbitrary time (from 0 to $\tau$) and finally divides by $\tau$, then it follows that:

$$m_h u_h^2 + x_h \zeta_h = m_h(x_{h\tau} u_{h\tau} - x_{ho} u_{ho})/\tau$$

where the values at time $\tau$ are characterised by lower index $\tau$ and the values at time zero by lower index o. By virtue of the stationary character of the motion $m_h(x_{h\tau} u_{h\tau} - x_{ho} u_{ho})$ must be smaller than $2m_h E\varepsilon/\tau$, given the value $E$ for the coordinates and $\varepsilon$ for the velocity components. If one allows the time of the entire motion $\tau$ to increase beyond any limit, then $2m_h E\varepsilon$ remains finite; hence the expression $2m_h E\varepsilon/\tau$ approaches zero. If one takes the mean value for a sufficiently long time, then:

$$m_h \langle u_h^2 \rangle + \langle x_h \zeta_h \rangle = 0$$

Similar equations are obtained for all coordinate directions and all material points. Adding, it follows that:

(143) $\quad \Sigma m_h c_h^2 + \Sigma \langle x_h \zeta_h + y_h \eta_h + z_h \zeta_h \rangle = 0$

½$\Sigma m_h c_h^2$ is the kinetic energy $L$ of the system. The expression $\Sigma \langle x_h \zeta_h + y_h \eta_h + z_h \zeta_h \rangle$ Clausius calls the virial of the forces acting on the system. Therefore, the above equation says that twice the time average of the kinetic energy is equal to the negative time average of the virial of the system during a very long time".

For a particle of mass $m$ in a circular orbit or radius $r$ around a mass $M$, the gravitational potential energy $V$ is $-GMm/r$ and the velocity is $(GM/r)^{1/2}$ so the kinetic energy $T$ is ½$GMm/r$. So $2T$ equals $-V$, just as explained for the virial theorem by Boltzmann and a result also obtained above holding constant the first derivative with respect to time of the moment of inertia, also physically action. This relationship holds equally well for changes of state for the kinetic energy and potential energy of particles. It follows that the total energy is $T + V = V/2$. So when gravitationally bound particles lose total energy (i.e. by radiation from the sustaining energy field as quanta), they speed up by falling, increasing their kinetic temperature. We see this as a catastrophic effect in the eventual collapse of stars. Succinctly, as commonly observed, compression causes heating.

It is conjectured that similar reversible or steady state processes for equilibration of kinetic and potential energy occur in the energy exchanges of molecules in the atmosphere with changes in height. Although clearly not in orbit around the Earth gravitationally bound molecules, partially suspended by the flow of heat in the atmosphere, can be regarded as in sub-orbital flight, reaching apogee at their maximum height dictated by their translational kinetic energy (Brown, 1968). Thus the differential change in potential energy for a molecule of *mgh* or $m(V_o^2 – V_n^2)$ transported from $h_o$ to $h_n$ should be equal to the sum of the decrease in kinetic energy $m(V_n^2 – V_o^2)/2$, a negative quantity indicating kinetic cooling *plus* an equivalent quantum of radiant energy ($hv$) absorbed by the gravity field. This now supports the slower moving particle higher in the gravitational field at a lower pressure. This is the specific sense in which Clausius' virial theorem is applied here.

*Estimating the lapse rate of temperature with height from the virial theorem*
It was concluded above from the virial theorem that the change in kinetic energy between the base level and height $h_n$ is also equal to half the negative increase in gravitational potential energy, or $–mgh_n/2$. This is an important conclusion as it allows the expected temperature change with height or lapse rate for molecules to be estimated simply as a function of their molecular weight. Comparing monatomic argon (mass 40) with helium (mass 4), for example, the rate of decline of temperature with height (if each gas was alone in the atmosphere) would be ten times greater at 15.86 K per km for argon and only 1.59 K per km for helium. This property will also differ for molecules with different internal energy modes and higher heat capacity.

A simple formula to obtain the theoretical temperature gradient for a quasi-equilibrium distribution is $mgh_n/2$ is equal to $nk\delta T/2$ yielding a lapse rate of $\delta T/h$ of $mg/nk$, where n indicates the degrees of freedom of action or kinetic motion able to contain heat, usually of $k/2$, although we will show quantum effects modify this freedom affecting n. For monatomic molecules like argon, n is 3, for diatomics like nitrogen and oxygen in their ground states for vibration it is 5 and for simple polyatomic molecules found in the atmosphere, it is 6 at moderate temperatures. However, in reality, the actual temperature gradient with altitude will be a cumulative variable, determined by the complex properties of other gases in the atmosphere, their mixing ratios and other local environmental factors such as temperature that may affect the vibrational heat capacity or quantum state. This is not a significant issue for the major diatomic gases in the atmosphere of Earth, but would be on Venus with its surface temperature in the vicinity of 740 K, with carbon dioxide the major gas.

In Table 1, virial-action lapse rates for the dry atmospheres of Earth, Venus and Mars are given, calculated using the formula above. Values for theoretical atmospheres on Earth assumed to have single gases are also given. These lapse values will be affected by the actual composition of the atmosphere. They could also be affected by changes of state such as condensation or other phase changes, as occurs for water on earth, releasing heat of condensation. The surface temperature can itself affect the heat capacity of greenhouse gases such as carbon dioxide as a quantum effect, therefore providing a feed-back effect on the gradient. The non-integral values for the degrees of freedom shown in Table 1 are values corrected for quantum effects of vibration, as used in Tables 5 and 6 following.

**Table 1:** Dry atmospheric virial gradients in temperature with altitude

| Planet and gas phase | Molecular weight $m$ Daltons | Surface gravity $g$ | Degrees of freedom n | $\delta T/h = mg/nk$ x$10^5$ K/cm |
|---|---|---|---|---|
| **Earth** | | 980.66 | | |
| Air ($N_2 + O_2 + A$) | 28.97 | " | 5.00 | 6.894* |
| Argon | 40.01 | " | 3.00 | 15.864 |
| Carbon dioxide | 44.01 | " | 5.41 | 9.650 |
| Water | 18.02 | " | 6.00 | 3.573 |
| **Venus** | | 887.0 | | |
| Carbon dioxide | 44.01 | " | 6.37 | 7.462 at 0 km |
| Carbon dioxide | 44.01 | " | 6.13 | 7.706 at 5 km |
| Carbon dioxide | 44.01 | " | 5.99 | 7.884 at 25 km |
| Carbon dioxide | 44.01 | " | 5.71 | 8.264 at 50 km |
| Carbon dioxide | 44.01 | " | 5.52 | 8.547 at 60 km |
| **Mars** | | 371.1 | | |
| Carbon dioxide | 44.01 | " | 5.11 | 3.866 at 0 km |
| Water | 18.02 | " | 6.11 | 1.324 at 0 km |

*The actual lapse rate observed in the Earth's atmosphere is about 6.5 K per km, probably reflecting heat released on condensation of water from humid air

*Gravity and thermodynamics of monatomic gases*

Based on the action method we developed for calculating entropy (Kennedy *et al*., 2014), we conclude the following equation describes differences in entropic energy $sT$ between a molecule transported from the base level $h_o$ to an altitude $h_n$ for a monatomic gas like argon. Considering the reversible exchange of two molecules at two altitudes varying both in gravitational potential and in entropic energy, we propose the following expression balancing gravity with thermodynamics.

$$mgh_n = kT_o\ln[e^{5/2}(@_{to}/\hbar)^3/z_t] - kT_n\ln[e^{5/2}(@_{tn}/\hbar)^3/z_t] - [3kT_o - 3kT_n] \qquad (5)$$

This unique relationship between gravity and thermodynamics is considered to describe a heat-work process for a particular species of molecule at the Earth's surface $h_o$ elevated to height $h_n$. Equation (4) would give a correct description of the molecular entropic energy $sT$ only for conditions of unchanged gravitational potential at $h_o$ and $h_n$. The differences between the entropic terms and kinetic terms $3kT_o - 3kT_n$ ($=mv_o^2 - mv_1^2$) at each altitude provide a balanced equation accounting for the increased work in gravitational potential energy $mgh$ (*vis viva* $= mV_o^2 - mV_1^2$) and the corresponding decrease in thermal sensible heat and work heat in the molecular system. Its derivation, considered as two simultaneous processes, requires application of the virial theorem as outlined above. The virial theorem, first enunciated mathematically by Laplace and Lagrange related to the calculus of variations and later by Clausius as indicated above, states that in a molecular ensemble a correspondence exists between the root-mean-square kinetic energy and half the potential energy, averaged over sufficient time. For equation (5), the question must be asked where the sources of energy in the $3k\delta T$ transferred with increased height are to be found. While one half is plainly kinetic energy, it is proposed that the second half must be a thermodynamic property of state, with heat released as chemical free energy increases with height acting as the source for the second half of the increased gravitational potential energy.

From application of the virial theorem to the atmosphere, we can conclude that half the change in potential $mgh_n/2$ (equal to the change in total or field energy) is equal to the difference in kinetic energy $(3kT_n - 3kT_o)/2$ – recalling that a negative change in kinetic energy occurs as molecular gravitational potential increases in the atmosphere. As a result, the temperature in the troposphere should fall at a nearly constant rate with height, subject only to small changes in *g*, or changes in quantum state and heat capacity of molecules with temperature. In the balanced equation (5) reversible work processes in gravity and thermodynamics or heat are expressed, with an ascending molecule A increasing its gravitational potential energy with a balancing descending molecule B elsewhere in the atmosphere subject to zero change in total energy. The total changes in thermal energy as heat or work are indicated by the change in entropy and temperature as $s_nT_n - s_oT_o$. In effect, a rising molecule uses the thermal energy associated with its entropy, both potential and kinetic absorption of heat, to do cooling gravitational work, increasing its free energy; a descending molecule uses its gravitational potential to generate heat as both kinetic and actinic emissions, increasing kinetic energy but reducing its free energy. Reversibility guarantees that air molecules at all altitudes tend to have the same capacity to do work.

We can rearrange equation (5) in the following form, to illustrate these effects.

$$mgh_n <= 3kT_o - 3kT_n = kT_o \ln e^{5/2}[(@_{to}/\hbar)^3/z_t] - kT_n \ln e^{5/2}[(@_{tn}/\hbar)^3/z_t]$$

Balancing the thermal terms,

$$mgh_n = kT_o \ln[(@_{to}/\hbar)^3/z_t] - kT_n \ln[(@_{tn}/\hbar)^3/z_t] - 1/2k(T_o - T_n) \qquad (6)$$

Given the ease of calculation of the terms of equation (7) at the surface we can solve for the change in number density and pressure with height by the following equation.

$$kT_n \ln[(@_{tn}/\hbar)^3/z_t] = kT_o \ln[(@_{to}/\hbar)^3/z_t] - 1/2k(T_o - T_n) - mgh_n \qquad (7)$$

Expressed as positive Gibbs energy terms we have,

$$mgh_n = kT_n \ln[(\hbar/@_{tn})^3 z_t] - kT_o \ln[(\hbar/@_{to})^3 z_t] - 1/2k(T_o - T_n) \qquad (8)$$

Expressed as Helmholtz energies, equation (8) becomes,

$$mgh_n = kT_n \ln[(\hbar/@_{tn})^3 z_t/e] - kT_o \ln[(\hbar/@_{to})^3 z_t/e] - 3/2k(T_o - T_n) \qquad (9)$$

These equations may initially be challenging to reconcile, because the two negative processes in energy on the right hand side of equation (9) translate to positive changes in potential energy on the left hand side. Furthermore, the first logarithmic free energy term is numerically negative, given the inversed relative action and the second at the surface is positive, but of greater magnitude since @$_{to}$ is less than @$_{tn}$. Overall, the difference between the two free energy terms is negative but, in terms of the quanta of energy released can be equated to gravitational work $mgh_n/2$. But for monatomic gases the virial theorem requires that $mgh_n/2$ must also equal the magnitude of $-3/2k(T_o - T_n)$. This property of a reversal in sign for kinetic energy is part of the nature of increasing potential energy. .

Here, translational action @$_t$ indicates a quantum level of $n\hbar$, where $\hbar$ is Planck's reduced quantum of action $h/2\pi$). This is a functional property of molecular momentum and radial separation, equal to $(3kTI_t)^{1/2}$ giving a quantum number $n_t$ by comparison to Planck's reduced quantum of action $\hbar$; here the moment of inertia for translational motion $I_t$ is equal to $mr_t^2$ the molecular mass multiplied by the square of the mean radial separation of the centre of mass of like molecules. The translational action @$_{to}$ at the base level is equal to $(3kT_oI_{to})^{1/2}$ whereas at altitude $h_n$ it will be greater of value $(3kT_nI_{tn})^{1/2}$, as temperature and pressure vary. Translational action plays a unique role in estimating entropy, acting as a functional surrogate for the effect of changes in both temperature and of volume or pressure.

The translational symmetry constant $z_t$ corrects the magnitude of the action to match the field energy required to sustain it; for the translation of ideal gases at 1 atm pressure, this factor was shown by us (Kennedy et al., 2015) to be of dimensionless magnitude around 10.23 (or 2.17 cubed) for all species of ideal gas molecules. This factor may be considered as preventing double-counting of molecules in translational molecular couples, with a residual correction of 8.5 per cent of the radial extension for statistical variation from a strictly cubic distribution of molecules. The $z_t$ factor has a similar effect as the symmetry factor $\sigma_r$ for rotation indicating the number of ways the atoms of a molecule can be oriented and remain identical morphologically. This rotational symmetry also reduces the field energy required to sustain the system.

Molecules in the atmosphere reversibly exchange kinetic energy plus Gibbs energy with gravitational energy. So molecule A ascending gains gravitational energy $mgh$ and molecule B descending gains the same kinetic energy as A loses ($mgh/2$) as well as losing the free energy that molecule A gains as it becomes colder. These logarithmic changes in action state involve the absorption or the emission of radiation, used to do gravitational work. In effect, the increase in free energy of an ascending molecule releases the quantum of heat needed for gravitational work, consistent with the virial theorem.

Comparing a molecule at the surface with one at altitude $h_n$, the increase in potential energy must equal the sum of the decrease in kinetic energy (expressed positively) and the change in free energy, which is also negative but expressed positively, given that $T_o$ is greater than $T_n$, while the ratio $\hbar/@_{to}$ is less than $\hbar/@_{tn}$ to a lesser degree, expressed logarithmically in equation (9).

An alternative means of expressing the above equations for a monatomic gas derived from equation (8) analogous to the barometric formula but with a non-isothermal atmosphere would be as shown in equation (10).

$$\{[(n_{tn})^3 e^{0.5}/z_t]^{T_n/T_o} / [(n_{to})^3 e^{0.5}/z_t]\} = e^{-mgh_n/kT_o} \qquad (10)$$

Note that here $n_{tn}$ and $n_{to}$ represent the cube root of the number density per cm$^3$. When compared with the Laplace formula, we see that factors for variation in kinetic temperature and translational action potential $\ln(@_{tn}/\hbar)^3$ are included. Equation (10) does bear some resemblance to the Laplace exponential relationship. However, there is no simple barometric relationship between pressure and altitude $h_n$, unless we assume that $T_n$ is equal to $T_o$ and the system is constrained to be isothermal, when equation (10) reduces to the Laplace formula.

In such an unrealistic case we would obtain, but with different action or quantum states,

$$[(n_{to})/(n_{tn})]^3 = e^{-mgh_n/kT_o}$$

Since the action depends on pressure or the number density only at constant temperature as shown in equation (1), this result suggests that Perrin's confirmation of the Laplace barometric equation on a microscope stage was possible as an approximation to isothermal conditions, dictated by the temperature of the suspending water. Perrin used an eyepiece blind with a pinhole to limit counts to a small number (less than 5) appearing in a single field of view. His counts were of monatomic particles in the small volume of fluid focused by the objective. However, despite the small number in focus, it can be concluded that his counts indicated the 3-dimensional number density rather than counts in the vertical dimension given by $n$.

As well as the change in action potential, to estimate the entropy change $\delta s$ per molecule between the surface and an altitude of $h_n$ we need to include the enthalpy change and not just the change in kinetic energy. The enthalpy for a monatomic gas includes both the kinetic energy $3/2kT$ as well as an additional $kT$ term to allow for the pressure-volume work that must be performed as a result of the Earth's atmospheric pressure at any altitude.

$$\begin{aligned}\delta s &= k\ln[e^{5/2}(@_{tn}/\hbar)^3/z_t] - k\ln[e^{5/2}(@_{to}/\hbar)^3/z_t] \\ &= k\ln[(@_{tn}/@_{to})^3]\end{aligned} \quad (11)$$

Because the translational action $@_n$ equal to $(3kT_nI_{tn})^{1/2}$ is a function of both temperature and radial separation, the logarithmic action increase with the decreased pressure at higher altitude is more than balanced by the linear decrease in temperature. Thus, the translational action and entropy increase with altitude although the entropic energy given by $sT$ decreases.

*Gibbs and Helmholtz energies*

Expressed per mole instead of the mean value per molecule, given $R$ is equal to $Nk$, where $N$ is Avogadro's number for the multiplicity of molecules in a mole, this becomes more familiar (Kennedy et al., 2015).

$$\begin{aligned}\delta s \delta T &= 3/2kT_o - 3/2kT_n + kT_n\ln[e(@_{tn}/\hbar)^3/z_t] - kT_o\ln[e(@_{to}/\hbar)^3/z_t] \\ &= (\Delta E - \Delta A)/N \\ &= (\Delta H - \Delta G)/N\end{aligned}$$

Here $\Delta E$ is the change in kinetic energy per mole, $N$ is Avogadro's number per mole, $\Delta A$ the change in Helmholtz free energy per mole, $\Delta H$ per mole is the change in enthalpy and $\Delta G$ the change in Gibbs energy. For this physical discussion we assume no effect of zero point vibrational energies on free energy, given no chemical reactions forming different bonds are involved. We recognize enthalpy as sensible heat, able to change the temperature and kinetic energy of molecules whether translational, rotational or vibrational. This explains how enthalpy can be measured using calorimeters and thermometers. But free energy and its inverse entropic energy is more subtle and difficult to measure, often appearing while heat that is absorbed as work is now "nowhere present", as stated by Clausius (1850) in explaining the reversible inter-conversion of heat and work, enabling caloric to be rejected as permanent heat.

Normally, these thermodynamic equations are written for reactions at constant temperature, but no such restriction is required here. Thus, $RT_n\ln[(@_{tn}/\hbar)^3/z_t] - RT_o\ln[(@_{to}/\hbar)^3/z_t]$ is the Gibbs energy change $\Delta G$ per mole for a change in temperature, lacking the extra $RT$ term required for changes in Helmholtz energy, expressed as $-RT\ln[e(@_t/\hbar)^3/z_t]$ or $RT\ln[(\hbar/@_t)^3 z_t/e]$. Obviously, both forms of free energy increase with increased altitude, since $RT_n\ln[e(@_{tn}/\hbar)^3/z_t] - RT_o\ln[e(@_{to}/\hbar)^3/z_t]$ must be negative, consistent with the spontaneous formation of the atmosphere, varying in pressure and temperature with altitude. Expressed positively, $G_o$ is equal to $RT_o\ln[(\hbar/@_{to})^3 z_t]$ – which always has a negative value – and $A_o$ is equal to $RT_o\ln[(\hbar/@_{to})^3 z_t/e]$ for a mole of a monatomic gas. So, as expected, $G_o$ is equal to $A_o + RT_o$ (see Glasstone (1951), Moore (1962)).

You can readily confirm this simple result as consistent with all the well known thermodynamic equalities given in texts, such as $\Delta A$ for *$\Delta E$ minus $T\Delta S$* and $\Delta G$ for *$\Delta H$ minus $T\Delta S$*. Keep in mind that in systems at equilibrium, changes in free energy $\Delta A$ or $\Delta G$ for physical or chemical reactions are extremely small or zero, with $\Delta E$ or $\Delta H$ equal to $T\Delta S$. Under equilibrium conditions, changes in translational entropy or free energy are equal to changes in internal entropy or free energy, so no work is possible.

In moving to higher gravitational potential, the quantum of radiation paying for this work process is abstracted from the thermal energy of the interior field (i.e. nearer the Earth's surface) and reassigned as gravitational energy to support the increased gravitational potential of the particle higher in the troposphere. Thus, $mgh_n$ can be separated into two equal segments for the decrease in kinetic energy and for the consequent equivalent decrease in the thermal or radiant energy $h\nu$; this absorption of radiant energy is given in equation (9) as the difference between the inverse free energy terms. Simply put, thermodynamic energy as a virial is exchangeable with gravitational energy. Taken together, the change in kinetic energy or temperature in the gravitational field with the change in free energy (or molecular thermal potential energy) constitutes the change in gravitational potential energy, justifying equations (5) and (9).

So from equation (9) for a monatomic gas we can write equation (12) a partial equation for half the change in gravitational potential per molecule, equal to the increase in Helmholtz energy with height (note the inverted action quotient), given that $@_{to}$ is less than $@_{tn}$ at higher altitude but $T_n$ is less than $T_o$, with greater effect.

$$mgh_n/2 = kT_n\ln[(\hbar/@_{tn})^3 z_t/e] - kT_o\ln[(\hbar/@_{to})^3 z_t/e]$$
$$= kT_o\ln[e(@_{to}/\hbar)^3/z_t] - kT_n\ln[e(@_{tn}/\hbar)^3/z_t] \qquad (12)$$

This equation can also be interpreted as showing that half of the gravitational work ($mgh_n/2$) is equivalent to the increase in Helmholtz energy per molecule. The inverting exponential terms in equation (12) account for the decrease in the pressure-volume function equivalent to $k\delta T$ for work potential with height as pressure declines. That is, the lower the temperature the more of the pressure-volume becomes available to do other work.

These conclusions from virial-action contrast strongly with the traditional view of an isothermal atmospheric system, where the total free energy at altitude $h_n$ is regarded as the sum of the Gibbs free energy and the total potential energy *mgh* (for example see Morowitz (1978)). These equilibrium state arguments lead to the Laplace barometric equation discussed earlier, whether derived classically, from statistical mechanics or from hydrostatics. In fact, despite their agreement, none of these derivations are true. The atmosphere is neither isothermal nor is it at equilibrium. The virial-action steady-state in the atmosphere can be considered as a quasi-equilibrium, entirely sustained by the heat flow. Without continuous though varying heat flow from the surface to space, the atmosphere would collapse. Indeed, the greater the flux of heat from the surface to space, the higher the atmosphere is elevated.

*Entropy and phase or action space*

Incidentally, freed from the need to include temperature except in defining the action, we can write the difference in mean entropy per molecule at the two altitudes as follows (Kennedy *et al.*, 2014).

$$\delta s = k\ln[e^{5/2}(@_{tn}/\hbar)^3/z_t] - k\ln[e^{5/2}(@_{to}/\hbar)^3/z_t]$$
$$= k\ln[(@_{tn}/@_{to})^3 = k\ln[(3kT_nI_n/3kT_oI_o)^{3/2} = 1.5k\ln[(T_nI_n/T_oI_o)] \qquad (14)$$

Given that the moment of inertia $I$ ($mr^2$) has spatial dimensions, with $r$ equal to $a/2$ where $a^3$ is the mean volume occupied by each molecular species, we can see how both temperature and volume can be included into translational action, $@_t$, equal to $(3kTI)^{1/2}$. Alternatively, changes in entropy can be related to changes in translational quantum numbers relative to $\hbar$. As shown in equation (14), the entropy change can be represented simply as a logarithmic function of the ratio of the cubic actions. In terms of averaged quantum states, we can express this entropy change approximately as a ratio of quantum numbers, as shown in equation (15).

$$\delta s = k\ln(n_{tn}/n_{to})^3 = 3k\ln(n_{tn}/n_{to}) \tag{15}$$

Note that the change in free energy includes the change in temperature between each altitude.

$$kT_n\ln[(n_{tn})^3/z_t] - kT_o\ln[(n_{to})^3/z_t] = -kT_o(\delta V^2/v_o^2) + k\delta T\ln K \tag{16}$$

where the constant $K$ is equal to $e^{5/2}$ and $\delta V^2$ is equal to $gh_n$.

To be consistent with the virial theorem, for a particle to be sustainably raised in a gravitational field from a position at $h_o$ to $h_n$, an amount of radiant (or thermal) energy equal to the resultant decrease in the kinetic energy of the molecule is required from the field. This is a reversible event that must be very frequent in the atmosphere. In a steady state equilibrium, every such thermal energy transfer to the gravitational field will be accompanied by a corresponding decrease in kinetic energy. So the correct barometric equation must account for the transfer of non-sensible work-heat to gravitational work, as well as the decrease in the heat supporting the kinetic energy in the elevated thermal system. This equivalence of gravitational and thermal potential energies proposed here is ultimately justified since functionally they must draw on the same energy field. The increase in free energy with height is half the increase in gravitational potential energy. The field energies of gravity and statistical thermodynamics exist separately only in text books.

*Equating gravitational and thermodynamic processes*
To illustrate the change in thermodynamic and gravitational parameters with altitude, results calculated using equation (7) for pure argon occurring alone at the Earth's surface temperature at 0.01 and 1.0 atmosphere surface pressure are shown in Table 2. It is assumed that the system is equilibrated at all altitudes although this may not be possible without sufficient rates of heat flow. The argon atmosphere could be considered as contained in a sealed, insulated chimney with a heat source of 288.15 K at its base equilibrated with the column of argon gas above. Sufficient argon is contained in the chimney to provide a weight giving a pressure equal to $1.013 \times 10^5$ pascals, or 1 atmosphere on Earth, or one-tenth this weight.

In the Earth's atmosphere, the one percent of argon is constrained to be at the same temperature as air, with a lapse rate near 6.5 K, rather than the 15.8 K per km altitude shown in Table 2 for argon alone. For both sets of data, note that the increase in gravitational potential energy $mgh_n$ shown in column 5 is equal to the difference between the increase in Gibbs energy and the decrease in enthalpy in column 6, in agreement with equation (8). Note that, assuming a steady state quasi-equilibrium, the column of argon will be at the Earth's black body temperature at 2.15 km from the surface, consistent with the greater lapse rate with argon. However, the pressure will have fallen to half its surface value of 1 atmosphere at around 805 metres from the surface, with a number density reaching half at around 850 metres. For an argon pressure of 0.0093 atm at the base, half this pressure value is reached even lower, at about 600 metres, with the halved density value occurring at 760 metres, reflecting the smaller number densities. However, as shown in Table 2, the thermodynamic changes with height are exactly the same irrespective of the surface pressure of the argon.

The data given for argon in Table 2 are theoretical, illustrating the plotting of atmospheric profiles using equation (7). Whether such a stable argon atmosphere is possible requires further study. Both profiles in Table 2 and those in all the subsequent tables for Earth, Venus and Mars imply equality of pressure, adjusted for height; the equality required is between that given by the weight of the atmosphere (as $p$ equal to $Mg$ per unit area, where $M$ is the mass of gas sustained above area $a^2$ and $g$ is gravity) with the thermodynamic pressure, equal to $kT$ per unit volume of $a^3$, the mean space occupied by each molecule of ideal gas.

$$p = Mg/a^2 = kT/a^3 \qquad (17)$$

Equation (17) implies that the pressure as weight per unit area due to a particular gas divided by the absolute temperature and the number density of each molecule will be approximately equal to Boltzmann's constant ($Mga^3/a^2 = Mga/T = k$). This requirement implies a close steady-state relationship between the atmospheric temperature at any height and the weight of the gas involved $Mg$ above the area $a^2$ generating the pressure from above at each height.

**Table 2:** Theoretical profile for an Earth-like planet for argon at 0.01 and 1.00 atm pressure

| Altitude (km) | Estimated temp. K | Estimate pressure x10$^{-3}$ Pascals | Estimated density x10$^{-17}$ /cm$^3$ | Gravitational potential energy $mgh_n$ x10$^{15}$ ergs per molecule | Decline in enthalpy $0.5k\delta T=\delta h$ x10$^{15}$ ergs per molecule | Increasing Gibbs energy $kT\ln(n_t)^3/z_t$ x10$^{13}$ ergs per molecule | Increase in Gibbs energy $-\delta kT\ln(n_t)^3/z_t$ x10$^{15}$ ergs per molecule |
|---|---|---|---|---|---|---|---|
| 0 | 288.15 | 9.419237 | 2.367675 | 0 | 0 | 8.2406263 | 0 |
| 1 | 272.33 | 3.008894 | 0.800271 | 6.552566 | -1.092094 | 8.1641802 | 7.64461 |
| 2 | 256.51 | 0.842474 | 0.237891 | 13.105133 | -2.184189 | 8.0877336 | 15.28927 |
| 3 | 240.69 | 0.201574 | 0.060660 | 19.657700 | -3.276283 | 8.0112870 | 22.93393 |
| 4 | 224.87 | 0.039897 | 0.012851 | 26.210266 | -4.368378 | 7.9348404 | 30.57859 |
| 5 | 209.05 | 0.006263 | 0.002170 | 32.762833 | -5.460472 | 7.8583938 | 38.22325 |
| 6 | 193.23 | 0.000737 | 0.000276 | 39.315399 | -6.552567 | 7.7819472 | 45.86791 |
| 7 | 177.41 | 0.000060 | 0.000025 | 45.867966 | -7.644661 | 7.7055006 | 53.51257 |
| 8 | 161.59 | 0.000003 | 0.000001 | 52.420533 | -8.736755 | 7.6290540 | 61.15723 |
| 9 | 145.77 | 0.000000 | 0.000000 | 59.020339 | -9.828850 | 7.5526074 | 68.80189 |
| 10 | 129.95 | 0.000000 | | 65.525666 | -10.920944 | 7.4761607 | 76.44656 |
| | | x10$^{-5}$ | x10$^{-19}$ | | | | |
| 0 | 288.15 | 10.12821 | 2.545887 | 0 | 0 | 6.3796980 | 0 |
| 1 | 272.33 | 4.24561 | 1.129198 | 6.552566 | -1.092094 | 6.3032514 | 7.64466 |
| 2 | 256.51 | 1.61311 | 0.455496 | 13.105133 | -2.184189 | 6.2268048 | 15.28932 |
| 3 | 240.69 | 0.54518 | 0.164063 | 19.657700 | -3.276283 | 6.1503582 | 22.93398 |
| 4 | 224.87 | 0.15820 | 0.051541 | 26.210266 | -4.368378 | 6.0739116 | 30.57864 |
| 5 | 209.05 | 0.03953 | 0.013698 | 32.762833 | -5.460472 | 5.9974650 | 38.22330 |
| 6 | 193.23 | 0.00079 | 0.002957 | 39.315399 | -6.552567 | 5.9210184 | 45.86796 |
| 7 | 177.41 | 0.00001 | 0.000491 | 45.867966 | -7.644661 | 5.8445718 | 53.51262 |
| 8 | 161.59 | 0.00000 | 0.000058 | 52.420533 | -8.736755 | 5.7681252 | 61.15728 |
| 9 | 145.77 | 0.00000 | 0.000004 | 59.020339 | -9.828850 | 5.6916785 | 68.80195 |
| 10 | 129.95 | 0.00000 | 0.000000 | 65.525666 | -10.920944 | 5.6152319 | 76.44661 |

The governing algorithm for monatomic molecules is that $mgh_{n-o} + 0.5k\delta T_{o-n} = \delta[kT_{o-n}\ln(n_t)^3{}_{o-n}/z_t]$ ;

The predicted pressure and density gradient for the theoretical atmosphere of argon in Table 2 is much sharper than that observed with air on Earth. Argon is constrained by its mixing temperature with the major gases, nitrogen and oxygen on Earth. It is interesting to consider the temperature profile in the atmosphere with solar warming at the surface, given the requirement that the Earth radiate with a mean black body temperature of 254 K. With no atmosphere, the mean temperature of 254 K would need to occur on the Earth's surface.

However, a role for heat conduction and resultant convection, even for an argon atmosphere, must be considered. Indeed, the thermal conductivity of argon at 288 K is 110% of that of carbon dioxide (Weast, 1964). In principle, a Carnot heat engine would function just as well with argon as a working fluid; we can propose that we would still find a heating effect above 254 K at the surface, even with no greenhouse gases present.

These equations for action or quantum states are consistent with the conclusion above that the total energy required to sustain all degrees of freedom in action in a chemical system at a given temperature $T_n$, is $S_n T_n$. It is reiterated here that this equilibration of thermal and gravitational action states should be regarded as the basis of the revised barometric formula.

$$Mgh_n + nR\delta T = \Delta S \Delta T = RT_o \ln[e^{5/2}(@_{to}/\hbar)^3 Q_e/z_t] - RT_n \ln[e^{5/2}(@_{tn}/\hbar)^3 Q_e/z_t]$$

From this equality, for a given surface temperature and atmospheric pressure of argon, it is possible to plot a steady state distribution of the atmosphere, sustained by the flow of heat. At the base level all the variables for computation can be taken as measured (temperature, pressure, molecular weight).

Does this mean that, based on the virial-action hypothesis, we can legitimately ask whether greenhouse gases capable of absorbing and radiating in the infrared are actually needed to obtain an inverted temperature gradient, heating the Earth's surface above its black body temperature?

*Polyatomic gases*
For molecules with more than one atom, including greenhouse gases, variation in rotational and vibrational action states must also be considered, particularly for the polyatomic molecules with three or more atoms. Both vibrational and rotational spectra of polyatomic molecules can be shown to contain fine structure as a result of absorption or emission of quanta of radio frequencies.

However, in the case of internal energy the order of higher and lower action states is reversed with altitude in the action barometric formula. We have shown above that the translational action or quantum number per molecule increases with altitude, even as temperature and translational velocity falls. But for rotation, the opposite change in quantum number occurs;

as molecules reach a higher altitude, they become colder and the rotational action states decline, decreasing their action and entropy, but increasing their Gibbs-Helmholtz energy. To stress the significance of this result, unlike the translational entropy the internal molecular entropy declines with altitude; conversely, the internal molecular Gibbs energy increases more rapidly with height as temperature falls. Note that for rotation, the Gibbs and Helmholtz energies are equal since no pressure-volume work can be done.

An electronic partition or multiplicity function, $Q_e$, corrects for electronic asymmetry in polyatomic molecules, such as diversity or asymmetry, given by parallel and non-parallel electron spins. This property is displayed by oxygen gas giving rise to three distinguishable sub-species and a $Q_e$ of 3. In effect each of these oxygen species has three times the action volume, increasing the action and entropy, but the total number of any one species in such an ensemble is reduced to one third as many. As a result, the heat capacity of the system estimated as $sT$ is increased compared to molecules with a single species like argon.

In the case of diatomic gases such as those predominant in the Earth's atmosphere nitrogen and oxygen the following equalities hold, analogous to equation (5) for monatomic gases.

$$\begin{aligned} mgh_n \; =& \; kT_o\ln[e^{5/2}(@_{to}/\hbar)^3 Q_e/z_t] - kT_n\ln[e^{5/2}(@_{tn}/\hbar)^3 Q_e/z_t] \; \text{(translation)} \\ &- kT_o\ln[e(@_{ro}/\hbar)^2/\sigma_r] + kT_n\ln[e(@_{rn}/\hbar)^2/\sigma_r] \; \text{(diatomic rotation)} \\ &- [5kT_o - 5kT_n] \end{aligned} \quad (18)$$

This is justified for a zero sum or balancing process on the same basis as for monatomic gases in equation (6), with entropic thermodynamic rearrangements on the right-hand-side, together with the $5kT_o - 5kT_n$ term for extraction of heat having an outcome as work equivalent to $mgh_n$. The thermodynamic process should be considered as occurring parallel to the gravitational process, as heat is converted to work. While it may appear that the same energy is being counted twice, this is not so. The kinetic terms on the right-hand side of equation (18) must be included to make the thermal adjustment required, matching the gravitational changes $mgh$. The outcome of this process is demonstrated in equation (19).

$$\begin{aligned} mgh_n \; =& \; kT_o\ln[(@_{to}/\hbar)^3 Q_e/z_t] - kT_n\ln[(@_{tn}/\hbar)^3 Q_e/z_t] \\ &- kT_o\ln[(@_{ro}/\hbar)^2/\sigma_r] + kT_n\ln[(@_{rn}/\hbar)^2/\sigma_r] \\ &- [7/2kT_o - 7/2kT_n] \end{aligned} \quad (19)$$

An even more complex virial-action version of the barometric equation similar to equation (10) could be derived from equation (19) if desired.

Given $@_{tn}$ is equal to $(3kTnI_n)^{1/2}$, a solution to the atmospheric profile can then be obtained as shown in equation (20).

$$kT_n \ln[(@_{tn}/\hbar)^3 Q_e/z_t]$$
$$= 3.5k(T_n - T_o) + kT_o \ln[(@_{to}/\hbar)^3 Q_e/z_t]$$
$$- kT_o \ln[(@_{ro}/\hbar)^2/\sigma_r] + kT_n \ln[(@_{rn}/\hbar)^2/\sigma_r] - mgh_n \quad (20)$$

Furthermore, by eliminating the kinetic energy of $2.5kT$ for nitrogen and oxygen we have the following relationship.

$$mgh_n/2 = kT_o \ln[(@_{to}/\hbar)^3 Q_e/z_t] - kT_n \ln[(@_{tn}/\hbar)^3 Q_e/z_t]$$
$$- kT_o \ln[(@_{ro}/\hbar)^2/\sigma_r] + kT_n \ln[(@_{rn}/\hbar)^2/\sigma_r] - 6k(T_o - T_n)$$

$$= kT_n \ln[(\hbar/@_{tn})^3 z_t/Q_e] - kT_o \ln[(\hbar/@_{to})^3 z_t/Q_e]$$
$$- kT_n \ln[(\hbar/@_{rn})^2 \sigma_r] + kT_o \ln[(\hbar/@_{ro})^2 \sigma_r] - 6k(T_o - T_n) \quad (21)$$

Although unphysical, equation (21) is mathematically correct if it is interpreted as showing that the difference between the changes in translational and rotational Gibbs energies at $h_o$ and $h_n$ is equivalent to the half the change in total potential energy, $mgh_n/2$ *plus* a consumption of heat equivalent to $6k(T_o - T_n)$. Once again, we have extracted the virial, as defined by Clausius and given statistical significance by Boltzmann and Gibbs.

In solving equation (20), the relative translational action $(@_{tn}/\hbar)$ at a definite altitude $h_n$ is computed, using known values for temperature and pressure at sea level $h_o$ to estimate all other functions. This is possible because rotational entropies and free energies for linear molecules like nitrogen and oxygen are a function of temperature only and not affected by changing pressure under Earth's ambient conditions. The translational moment of inertia ($I_t = mr_t^2$) is then calculated and the number density and pressure at $h_n$ can easily be found. It should be observed that the kinetic energy factor of $2.5k\delta T$ for a linear diatomic gas like nitrogen (N$_2$) or oxygen (O$_2$) *must* equate with half the change in gravitational potential energy (*mgh*/2) to comply with the virial theorem, as discussed above.

Thus, as a logical extension for polyatomic molecules, the following general equation including vibration needed for greenhouse molecules, indicates the result of the interaction between gravitational and thermodynamic potential. Note that, as in equation (18), the right hand side of equation (18) expresses differences in entropic energy,

$$mgh_n + n(kT_o - kT_n) = kT_o\ln[e^{5/2}(@_{to}/\hbar)^3 Q_e/z_t] - kT_n\ln[e^{5/2}(@_{tn}/\hbar)^3 Q_e/z_t]$$
(translation)
$$- kT_o\ln[e(@_{ro}/\hbar)^2/\sigma_r] + kT_n\ln[e(@_{rn}/\hbar)^2/\sigma_r]$$
(rotation of linear molecules)

OR

$$- kT_o\ln[\pi^{1/2}e^{3/2}(@_{Ao}@_{Bo}@_{Co}/\hbar^3)/\sigma_r] + kT_n\ln[\pi^{1/2}e^{3/2}(@_{An}@_{Bn}@_{Cn}/\hbar^3)/\sigma_r]$$
(nonlinear rotation)
$$+ \Sigma\{kT_o[x_o/(e^{x_o} - 1) - \ln(1 - e^{-x_o})] - kT_n [x_h/(e^{x_h} - 1) - \ln(1 - e^{-x_h})]\}$$
(vibration)     (22)

In equation (22), n is the degrees of freedom of kinetic motion requiring heat determining the lapse rate in temperature, which is 5 for linear molecules like nitrogen and oxygen at the moderate temperatures of the Earth's atmosphere. For estimating vibrational entropy, *x* is equal to the ratio *hv/kT* where *v* is the vibrational frequency expressed in Hz. Vibrational entropy must be the sum shown in equation (22) using all resonant wavelengths (Moore, 1962), including any redundancy (Kennedy et al., 2015). In the case of significant vibrational entropy for carbon dioxide as on the surface of Venus, the virial theorem requires that equation (20) must have a function $mgh_n + (5 + n_{vib})(kT_o - kT_n)$ on the left hand side, since the factor n = (5 + $n_{vib}$) determines its lapse rate, according to the formula $\delta T/h = mg/nk$ given earlier. Obviously, for carbon dioxide on Venus, the lapse rate will vary with altitude, being lower at low altitudes where the temperature is higher. So the vibrational heat capacity must be estimated as well as the differences in the *sT* entropy values on the right hand side of the equation. Fortunately, this can be easily achieved for vibration using the relationship given in Walter Moore's text book, discussed in Kennedy *et al*. (2014).

$$\partial E/\partial T = C_V = Rx^2/[2(\cosh x - 1)]$$

Unlike translation and rotation, the virial for a vibrating harmonic oscillator involves equality of the mean values of the kinetic energy and the potential energy, with no change of sign. They increase or decrease in reversible transitions together and the heat capacity $C_v$ as measured experimentally must involve equal amounts of both kinetic and potential energy. Therefore the change in kinetic degree of freedom of motion must involve $\delta C_v/2$. Consequently, $\delta n_{vib}$ is equal to $\delta C_v/2$. This surprising conclusion was achieved by personal communication (Jacob Linder, Norwegian University of Science and Technology, 2014). As a result, $n_{vib}$ must be equal to $Rx^2/[4(\cosh x - 1)]$ for each oscillation in the molecule.

Equation (20) can be used to solve for changes in translational action (@$_{tn}$) and pressure with altitude, since free energies at the base level can be calculated from known properties such as temperature and pressure. Internal energies are a function only of temperature, so all the rotational and vibrational terms at any height can also be calculated when temperature is known. For polyatomic gases, the change in total kinetic energy that is a function of the molecular heat capacity is still equal to half the change in gravitational potential energy ($mgh/2$); but the latent heat factor appearing as negative free energy (also equal with argon to half the change in potential energy as shown in equation (3)) is now considerably more complex. It involves a relationship in which changes in translational and vibrational energies and changes in rotational energy make separate but opposing contributions, as indicated in equations (20) and (22).

It is essential to comprehend the basis for inverted signs with altitude in equations (19) and (20) for the rotational action compared to translational action. As altitude increases, all free energies increase as temperature falls, releasing heat as quanta. While the changes in action and free energy for translation and vibration bear the same sign as that for increased gravitational potential, for rotation the decrease in action and free energy operates with an inverted sign. This is required because the quantum levels for translation and rotation vary in opposite directions with increased altitude. The translational action state given by the ratio @$_t/\hbar$ must increase as pressure falls, to conform with the ideal gas law, whilst rotational action @$_r/\hbar$ decreases as temperature falls, releasing energy as quanta needed to provide gravitational energy.

The decline in rotational quantum number that is obvious from the analysis of equation (20) corresponds with an increase in translational quantum number. It can be considered that as the molecules execute Brownian rotation more slowly as temperature falls and less action, they travel on longer translational radii having translational trajectories with equally more action. This is a Lagrangian variational phenomenon of the system as it achieves least action. This requires that the mean size of rotational quanta increase with altitude while those for translation decrease, as the translational quantum level increases.

In chemical reactions at constant pressure or volume, decreases in internal energy or enthalpy and internal entropy between reactants and products are compensated by increases in translational energy and entropy and *vice versa* as the quasi-equilibrium is reached. For equilibria between gaseous isomers with different translational and rotational moments of inertia, a position of equilibrium is reached where $T\Delta S$ is equal to $\Delta H$ *and* $\Delta G$ is zero. Thus, in readjusting from a standard state of unit pressure of both isomers, the $\Delta G^o$ for the reaction will decline to zero. At constant temperature where the internal free energies (rotation, vibration, electronic) do not vary, the translational energy enjoys a special relationship with chemical potential as the equilibrium constant can be expressed as $\Delta G^o = -RT\ln K_{eq}$.

Similar thermodynamic processes occur physically with changes in altitude and temperature even without chemical reaction, involving gravitational work. Internal molecular entropy declines with the colder temperature, in contrast to the increasing translational entropy at increased elevation and decreased pressure but greater translational action. The declining temperature with height still ensures that translational Gibbs energy increases despite the expansion (see equations (10) and (15)) and pressure-volume work, as the thermal energy transitions to colder wavelengths.

In Table 3, the thermodynamic properties of air with properties averaged for nitrogen and oxygen with root-mean-square velocities characteristic of a surface temperature of 288.15 K are shown. These data were calculated using numerical computation, as outlined in Figure 1, for each height. A comparison between data generated by the action barometric model and actual field observations collated by the United States Air Force (Anderson *et al*., 1986) is also made in Table 3.

**Table 3:** Thermodynamic profile on Earth estimated for atmospheric molecules

| Altitude (km) | Temp K | Estimate pressure x$10^{-5}$ Pascals | Estimate of density x$10^{-19}$ per cm$^3$ | Solved translational Gibbs energy $kT\ln(n_t)^3 Q_e/z_t$ x$10^{13}$ergs per molecule | Estimated from $T$ as increasing rotational Gibbs energy $kT\ln[(n_r)^2/\sigma_r]$ x$10^{13}$ergs per molecule | USAF Temp K | USAF Model 6 Pressure |
|---|---|---|---|---|---|---|---|
| 0 | 288.2 | 1.01282 | 2.545887 | 6.3252617 | 1.5965459 | 288.2 | 1.0130 |
| 1 | 281.3 | 0.89895 | 2.315004 | 6.1968176 | 1.5489674 | 281.7 | 0.8988 |
| 2 | 274.4 | 0.79384 | 2.095675 | 6.0686066 | 1.5016219 | 275.2 | 0.7950 |
| 3 | 267.5 | 0.69722 | 1.888033 | 5.9406345 | 1.4545154 | 266.7 | 0.7012 |
| 4 | 260.6 | 0.60880 | 1.692180 | 5.8129076 | 1.4076540 | 262.2 | 0.6166 |
| 5 | 253.7 | 0.52826 | 1.508188 | 5.6854323 | 1.3610441 | 255.7 | 0.5405 |
| 6 | 246.8 | 0.45527 | 1.336097 | 5.5582154 | 1.3146928 | 249.2 | 0.4722 |
| 7 | 239.9 | 0.38496 | 1.175904 | 5.4312642 | 1.2686071 | 242.7 | 0.4111 |
| 8 | 233.0 | 0.33059 | 1.027566 | 5.3045863 | 1.2227945 | 236.2 | 0.3565 |
| 9 | 226.1 | 0.27817 | 0.890993 | 5.1781896 | 1.1772636 | 229.7 | 0.3080 |
| 10 | 219.2 | 0.23187 | 0.766044 | 5.0520831 | 1.1320224 | 223.3 | 0.2650 |

The data set was computed using equation (18), assuming a base temperature of 288.15 K and the Earth's average pressure at sea level, with gravity ($g$) 980.66. A standard diatomic molecule of mass 29, with bond length 1.13x$10^{-8}$ cm, a rotational symmetry of 2 and a $Q_e$ value of 1.41 to account for oxygen's spin multiplicity of 3 was used. However, the same data could have been produced using the respective proportions of nitrogen and oxygen. Minor gases including argon were ignored in the calculation. The above plot was made by solving the equation (20) $kT_n\ln[(@_{tn}/\hbar)^3 Q_e/z_t] = 3.5k(T_n - T_o)$ + $kT_o\ln[(@_{to}/\hbar)^3 Q_e/z_t] - kT_o\ln[(@_{ro}/\hbar)^2/\sigma_r] + kT_n\ln[(@_{rn}/\hbar)^2/\sigma_r] - mgh_n$.

Do keep in mind that Gibbs energy (*g*) or Helmholtz energy (*a*) indicating the work potential actually expresses the inverse of the field energy content; it can also be thought of as indicating a potential for increased action, allowing absorption of quanta. The corresponding decrease in the translational and rotational kinetic energy of the molecular particles is included in the changing enthalpy term in column 3 of Table 4. At ambient temperatures near 288 K, there is no significant need to include vibrational entropy and the corresponding Gibbs/Helmholtz energy changes with nitrogen and oxygen gases as these vibrational free energies vary only very marginally with height.

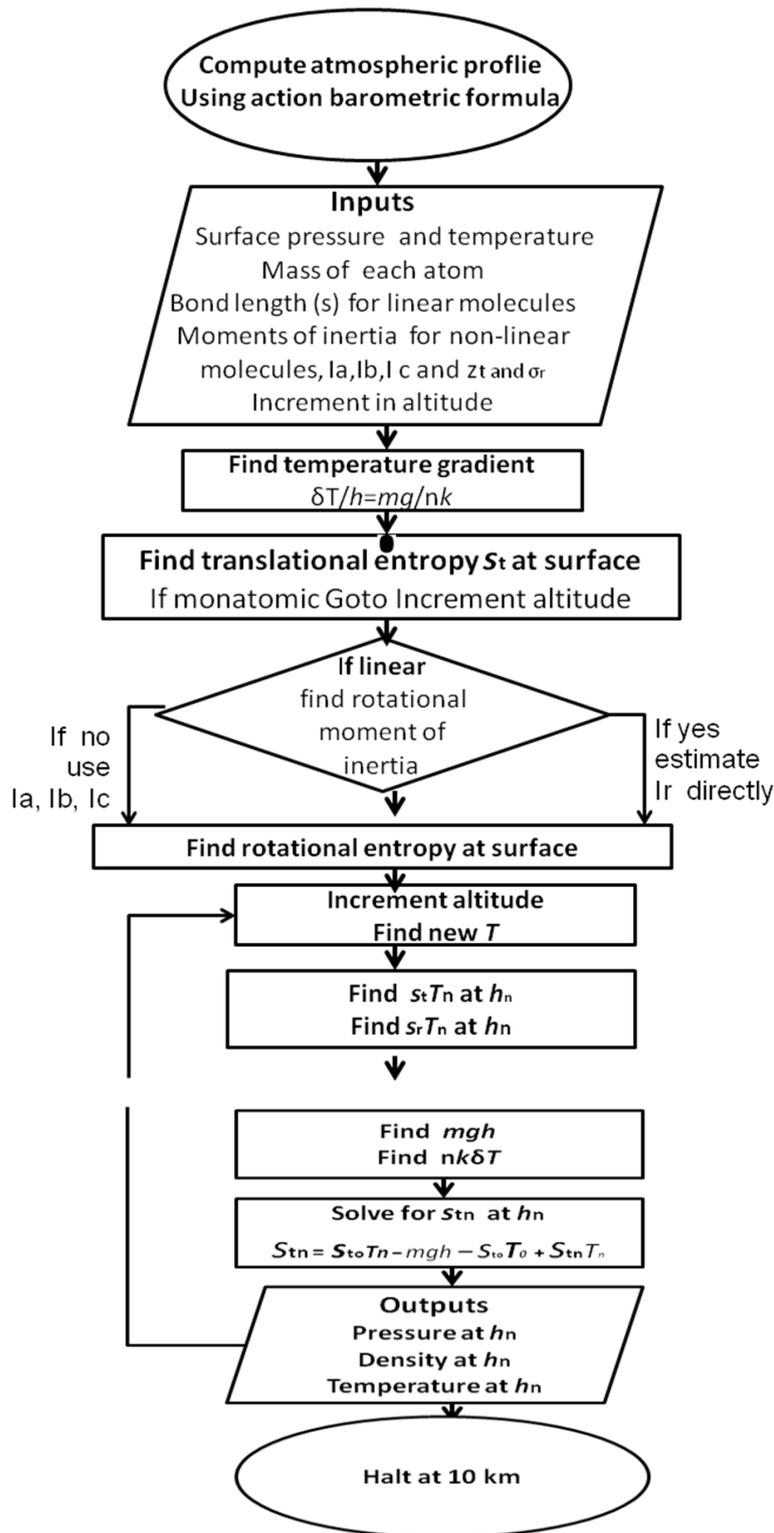

**Figure 1:** flow sheet for computation of atmospheric profile using the virial theorem to obtain translational entropy, pressure and temperature at altitude $h_n$. A fully annotated version of this programme is available from the corresponding author. The new data were calculated as a solution to equation (19), effectively producing the cumulative increase in translational Gibbs energy (i.e. the decrease in its inverse) shown in column 5. The computer model used estimated changes in temperature, pressure and density as well as thermodynamic and gravitational properties using inputs of the surface temperature and pressure alone.

**Table 4:** Changes in gravitational and thermodynamic properties with height on Earth

| Altitude (km) | Estimated gravitational potential energy $mgh_n$ x$10^{15}$ erg per molecule | Estimated decreasing enthalpy $3.5k\delta T = \delta h$ x$10^{15}$ ergs per molecule | Estimated rotational Gibbs energy $\delta[kT\ln(n_r)^2/\sigma_r]$ $= -\delta g_r$ x$10^{15}$ ergs per molecule | Solved translational Gibbs energy $\delta[kT\ln(n_t)^3/z_t]$ $= -\delta g_t$ x$10^{15}$ ergs per molecule | Solved cumulative translational Gibbs energy |
|---|---|---|---|---|---|
| 0 | 0 | | | | |
| 1 | 4.75679 | -3.32976 | -4.75783 | -12.84441 | -12.84438 |
| 2 | 9.51359 | -6.65951 | -9.49241 | -12.82111 | -25.66551 |
| 3 | 14.27038 | -9.98927 | -14.20306 | -12.79721 | -38.46271 |
| 4 | 19.02718 | -13.31903 | -18.88920 | -12.77269 | -51.23541 |
| 5 | 23.78397 | -16.64878 | -23.55018 | -12.74753 | -63.98294 |
| 6 | 28.54077 | -19.97854 | -28.18532 | -12.72169 | -76.70463 |
| 7 | 33.29756 | -23.30829 | -32.79389 | -12.69512 | -89.39975 |
| 8 | 38.05436 | -26.63805 | -37.37514 | -12.66779 | -102.06754 |
| 9 | 42.81115 | -29.96781 | -41.92824 | -12.63967 | -114.70720 |
| 10 | 47.56795 | -33.29756 | -46.45235 | -12.61065 | -127.31787 |

The governing algorithm for diatomic or other linear molecules is that
$mgh_{n\text{-}o} + 3.5k\delta T_{o\text{-}n} = \delta[kT_{o\text{-}n}\ln(n_t)^3_{o\text{-}n}/z_t] - \delta[kT_{o\text{-}n}\ln(n_r)^2_{o\text{-}n}/\sigma_r] = \delta g_t - \delta g_r$;

Minor differences in the two profiles obtained by action theory and the US Air Force Model 6 are of no concern (see Figure 2). The mean observations given as USAF Model 6 (Anderson et al., 1986) in Table 3 represent actual data collated and averaged from large numbers of measurements at different locations. In contrast, the profile calculated from action thermodynamics is specific to one set of uniform atmospheric surface conditions at one location. The differences correspond to slightly different temperature profiles but these match the differences in pressure in any case. It would be possible to adjust the action model to achieve a better match to Model 6, but this would be artificial and little or no advantage would be gained. Furthermore, the profile was estimated assuming ideal gas behavior. A subroutine to correct for non-ideal behavior can easily be included in the action model.

In Table 4, the equality predicted in equation (20) for diatomic ideal gases on Earth is illustrated, with the cumulative translational Gibbs energy obtained as a solution shown in column 5 being given by the difference between the increasing gravitational potential energy (column 2) and the decreasing enthalpy (column 3) plus the increasing rotational free energy (column 4).

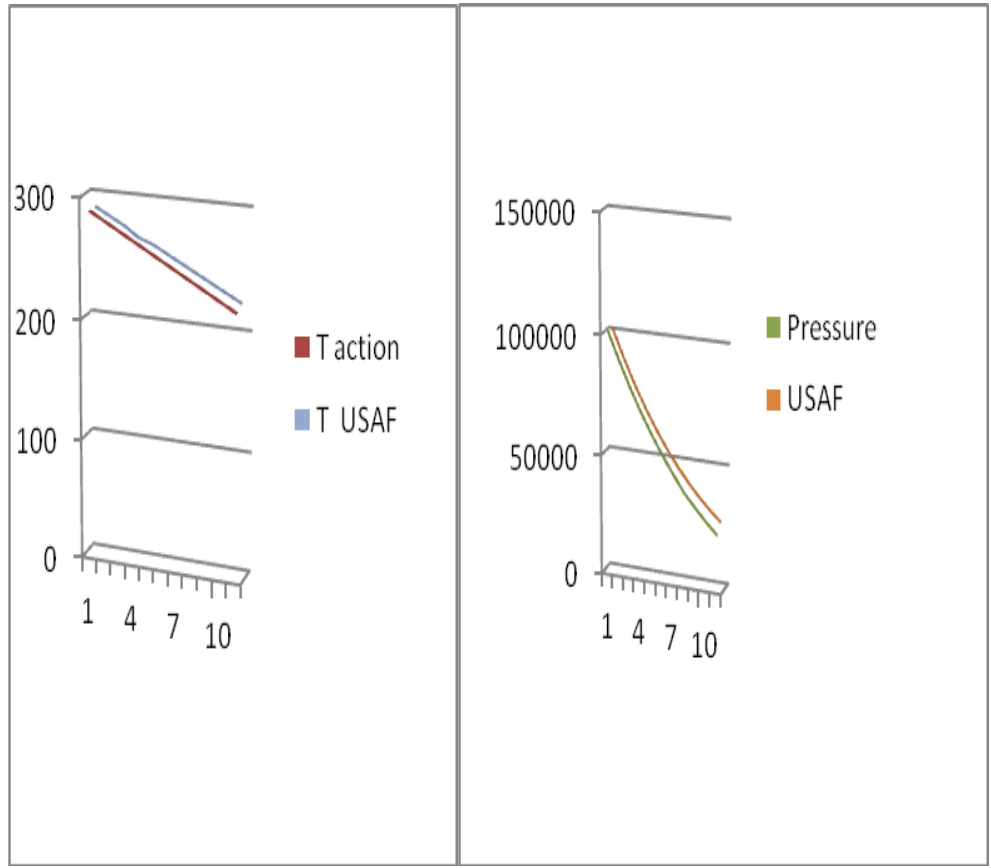

**Figure 2:** Comparison of temperature and pressure profiles from action thermodynamics and USAF Model 6 observations [17]. X-axis is altitude in km and Y-axes are degrees Kelvin and bars respectively. Temperatures fall linearly while pressure falls exponentially.

Thus gases in the atmosphere show increased translational action and entropy with altitude corresponding to decreased pressure, despite the decrease in temperature. Their cubic translational action – a function of $(3kTI_t)^{3/2}$ – is more responsive to decreased pressure and increased radial separation affecting their inertia ($I_t$) than to their temperature ($T$). Despite this increase in action, the sustaining thermal field energy requirement decreases and the Gibbs and Helmholtz energies increase because the effect of the linear decrease in temperature with altitude exceeds the effect of the logarithmic value of the increase in translational action and increased entropy. The higher translational quantum state at higher altitude involves energy transitions of lower magnitude, with quanta of longer wavelength (see Table 4). Overall, less thermal energy is required for maintaining each molecule in a higher quantum state at the lower temperature. This difference in the thermal energy requirement with height releases energy to pay for part or all of the work of greater gravitational field energy, indicated by $mgh/2$.

Rotational action and entropy are also subject to changes required for action in linear molecules ($CO_2$, $N_2$, $O_2$) of $2kTI_r$ or in 3-dimensional molecules ($H_2O$, $CH_4$, $N_2O$) of $(2kTI_r)^{3/2}$ but the rotational moment of inertia is regarded here as fixed because bond lengths respond slowly to changes in temperature– except at much higher temperatures than 300 K, where vibration also becomes more significant. As a result, by contrast to translation, the action and entropy for rotation and vibration decrease linearly with temperature or according to $T$ ($CO_2$, $N_2$, $O_2$) or $T^{3/2}$ ($H_2O$, $CH_4$, $N_2O$, etc.). Consequently, internal entropy decreases with altitude and, conversely, the free energy increases, between the base level of zero height $h_o$ and $h_n$. The internal energy decreases by the product of a linear factor with height according to temperature as a well as a logarithmic factor (either $\ln T$ or $\ln T^{3/2}$) – i.e. by $\delta T \delta \ln T$.

An illustration of the operation of the virial-action hypothesis for a global steady-state atmosphere is shown in Figure 3. This is based on the results of Table 3 for an Earth atmosphere of mean molecular mass 29, weighted for the relative concentrations of $N_2$ and $O_2$ regarding mean bond length, rotational symmetry and the electronic multiplicity factor ($Q_e$ = 1.41 rather than 3 needed for oxygen). More than half the atmosphere is contained within 5.9 km of altitude in the troposphere, less than 0.1% of the distance to the Earth's centre – or in a shell less than one-millionth of the earth's volume, a fragile zone indeed on which we utterly depend. A future version of this virial-action model will include the heat and heat-work exchanges involved in advection and convection of the atmosphere.

Tables 3 and 4 gave approximate results, based on the formula using the median properties of nitrogen and oxygen, the major gases in the Earth's atmosphere. In Tables 5 and 6, solutions for the atmospheres of Venus and Mars, which are predomissnantly carbon dioxide, are given. The results in these profiles correspond closely to the reported observations made on these planets, showing the strong "tidal" effect observed on Mars as a function of the surface temperature (Table 6). An anomaly observed was that, unlike translational and rotational entropies for the diatomic gases (equations (7) and (9)), where entropic energy includes components that can be contributed to both kinetic and free energies, the classical formula for vibrational entropy given in texts (e.g. Moore, 1962) apparently does not include an equivalent term for kinetic energy ($kT\ln e^{n/2}$). As a result, the formula needed to calculate the translational entropic energy $s_t T$ ($kT_o\ln[e^{5/2}(@_{to}/\hbar)^3 Q_e/z_t]$) at each altitude does not require the term giving twice the kinetic energy calculated from the heat capacity ($C_{vib}$), as was required for translation and rotation. Shown in Tables 5 and 6, the entropic energy term ($sT$) is frequently exceeded by the enthalpy term $c_{vn}T_n$, particularly at lower temperatures as on Mars. However, the marked difference between translational and rotational heat capacities compared to vibrational heat capacity regarding potential energy dealt with earlier (equation (22)) explains this apparent anomaly (see also equation (25) below.

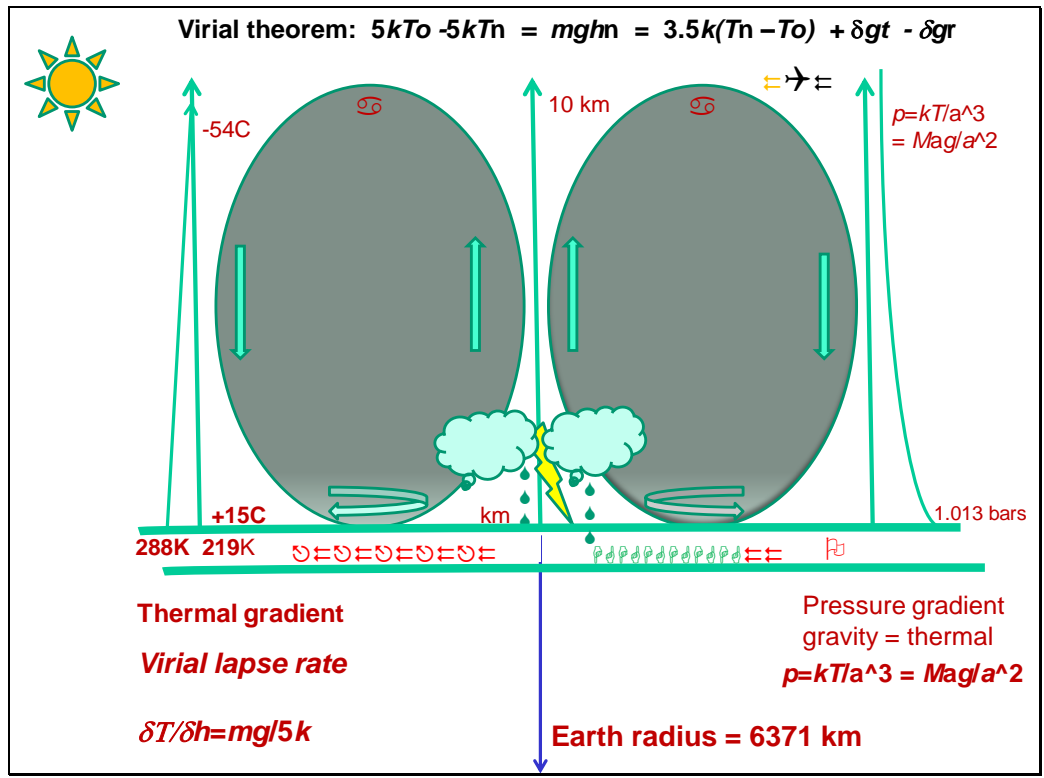

Figure 3: The virial-action thermodynamic model of the Earth's atmosphere. Linear, slightly curvilinear and logarithmic gradients are shown for temperature, variation in Gibbs energy and pressure respectively. The virial theorem justifies a symmetrical consumption of heat from the kinetic energy ($5/2kT$) and Gibbs energy fields, mainly from $N_2$ and $O_2$, each equivalent to $mgh/2$ of gravitational work. At all heights, it is essential that pressure is related approximately to temperature and number density or volume by the ideal gas equation, $p = kT/a^3$, where $a^3$ represents the mean volume available to each colliding molecule.

Then we can write equation (23) as indicating a process involving extraction of thermodynamic heat equivalent to $mgh_n$ from the separate entities relating gravitational and thermodynamic transitions.

$$mgh_n + 5(kT_o - kT_n) + \delta c_v \delta T = kT_o \ln[e^{5/2}(@_{to}/\hbar)^3 Q_e/z_t] - kT_n \ln[e^{5/2}(@_{tn}/\hbar)^3 Q_e/z_t]$$
$$- kT_o \ln[e(@_{ro}/\hbar)^2/\sigma_r] + kT_n \ln[e(@_{rn}/\hbar)^2/\sigma_r]$$
$$\text{OR}$$

$$- kT_o\ln[\pi^{1/2}e^{3/2}(@_{Ao}@_{Bo}@_{Co}/\hbar^3)/\sigma_r] + kT_n\ln[\pi^{1/2}e^{3/2}(@_{An}@_{Bn}@_{Cn}/\hbar^3)/\sigma_r]$$
$$+ \Sigma\{kT_o[x_o/(e^{x_o} - 1) - \ln(1 - e^{-x_o})] - kT_n[x_h/(e^{x_h} - 1) - \ln(1 - e^{-x_h})]\} \quad (23)$$

where $C_V$ is equal to $Rx^2/[2(\cosh x - 1)]$ per mole or $c_v$ is equal to $kx^2/[2(\cosh x - 1)]$ per molecule.

Once again we may write, by extracting the thermal energy equivalent to half the change in potential energy $mgh/2$ or $5/2(kT_o - kT_n) + \delta c_v \delta T/2$; only the kinetic half of the vibrational heat capacity is required.

$$mgh_n/2 = kT_o\ln[e(@_{to}/\hbar)^3 Q_e/z_t] - kT_n\ln[e(@_{tn}/\hbar)^3 Q_e/z_t] - kT_o\ln[(@_{ro}/\hbar)^2/\sigma_r] +$$
$$kT_n\ln[(@_{rn}/\hbar)^2/\sigma_r] + \Sigma\{kT_o[x_o/(e^{x_o} - 1) - \ln(1 - e^{-x_o})] - kT_n[x_h/(e^{x_h} - 1) - \ln(1 - e^{-x_h})]\} \quad (24)$$

Just as before in equations (12) or (20) for monatomic and diatomic gases like argon, nitrogen and oxygen, we interpret this result as meaning that half the change in molecular potential energy is equal to changes in Helmholtz energy in the transition from altitude $h_o$ to altitude $h_n$, providing the virial.

Thus, the carbon dioxide profiles for Venus and Mars (see Tables 5 and 6) were calculated employing equation (25), obtained by re-arranging equation (23).

$$kT_n\ln[(@_{tn}/\hbar)^3 Q_e/z_t]$$
$$= 3.5k(T_n - T_o) + kT_o\ln[(@_{to}/\hbar)^3 Q_e/z_t]$$
$$- kT_o\ln[(@_{ro}/\hbar)^2/\sigma_r] + kT_n\ln[(@_{rn}/\hbar)^2/\sigma_r] - mgh_n$$
$$+ \Sigma\{kT_o[x_o/(e^{x_o} - 1) - \ln(1 - e^{-x_o})] - kT_n[x_h/(e^{x_h} - 1) - \ln(1 - e^{-x_h})]\}$$
$$+ \Sigma\{kT_o x_o^2/[4(\cosh x_o - 1)] - kT_n x_n^2/[4(\cosh x_n - 1)]\} \quad (25)$$

In Table 6, the temperature and pressure data similar to that calculated in Table 5 are compared with measurements of the Venus atmosphere obtained with the Magellan spacecraft (Jenkins *et al.*, 1994). The data in Table 5 are a first pass product from application of the formula given in equation (25) and no adjustments have been made to obtain better agreement with the Magellan results, obtained at an latitude of 67 degrees north. As Table 6 shows, using the surface temperature measured by the Magellan lander, there is very good agreement in the temperature profiles even up to 60 km altitude. This result is despite the atmosphere being mainly carbon dioxide, well known to be a non-ideal gas suggesting that corrections for non-ideality is required, perhaps by applying the van der Waals equation or the virial equation. However, at higher temperatures, non-ideal gases have better agreement with the ideal gas law. The Magellan data in column 8 do show significant divergences with altitude from the gas law, given the significant differences. Either the observations lack precision, or other physical factors such as non-ideality were operating.

**Table 5:** Calculated thermodynamic profiles on Venus for a 96.5% $CO_2$ atmosphere

| Altitude (km) | Temp estimate K | Pressure estimate x$10^{-6}$ Pascals | Density estimate x$10^{-20}$ per cm$^3$ | Translational Gibbs energy solution $kT\ln(n_t)^3Q_e/z_t$ x$10^{13}$ergs per molecule | Rotational Gibbs energy estimate from $T$ $kT\ln[(n_r)^2/\sigma_r]$ x$10^{13}$ergs per molecule | Vibrational entropic energy x$10^{13}$ergs per molecule | Vibrational thermal energy from $T$ x$10^{13}$ergs per Molecule |
|---|---|---|---|---|---|---|---|
| **Venus** | | | | | | | |
| 0 | 740.0 | 8.9918 | 8.8011 | 14.357464 | 6.738396 | 1.979265 | 1.691308 |
| 5 | 702.4 | 4.6428 | 4.7879 | 14.141371 | 6.344927 | 1.752875 | 1.231999 |
| 10 | 664.2 | 3.2900 | 3.5879 | 13.560185 | 5.948694 | 1.535053 | 1.123855 |
| 15 | 626.1 | 2.2626 | 2.6204 | 12.961938 | 5.549468 | 1.326512 | 1.017659 |
| 20 | 586.0 | 1.5015 | 1.8559 | 12.345019 | 5.147028 | 1.128077 | 0.913709 |
| 25 | 545.8 | 0.9542 | 1.2662 | 11.707705 | 4.741143 | 0.940697 | 0.812242 |
| 30 | 504.9 | 0.5746 | 0.8242 | 11.048112 | 4.331498 | 0.765437 | 0.713371 |
| 35 | 465.3 | 0.3226 | 0.5046 | 10.364059 | 3.917543 | 0.603472 | 0.616991 |
| 40 | 420.2 | 0.1648 | 0.2840 | 9.652687 | 3.498189 | 0.456070 | 0.522684 |
| 45 | 375.9 | 0.0734 | 0.1414 | 8.756594 | 2.984649 | 0.300395 | 0.411112 |
| 50 | 329.5 | 0.0263 | 0.0578 | 8.127117 | 2.632570 | 0.210825 | 0.336836 |
| 55 | 280.1 | 0.0065 | 0.0168 | 7.290891 | 2.174624 | 0.117102 | 0.243483 |
| 60 | 225.9 | 0.0007 | 0.0094 | 6.375712 | 1.686714 | 0.060019 | 0.170513 |
| | | | | | | | |

Profiles in Table 5 and 6 were plotted using data from NASA (2014) by solving equation (25)
$kT_n\ln[(@_{tn}/\hbar)^3Q_e/z_t] = 3.5k(T_n - T_o) + kT_o\ln[(@_{to}/\hbar)^3Q_e/z_t] - kT_o\ln[(@_{ro}/\hbar)^2/\sigma_r] + kT_n\ln[(@_{rn}/\hbar)^2/\sigma_r] - mgh_n + \Sigma\{kT_o[x_o/(e^{xo} - 1) - \ln(1 - e^{-xo})] - kT_n[x_h/(e^{xh} - 1) - \ln(1 - e^{-xh})]\} + kT_ox_0^2/[4(\cosh x_o - 1)] - RT_nx_n^2/[4(\cosh x_n - 1)]$.

Despite this, it can be concluded from Table 6 that as, a test of the virial-action temperature lapse rate, the observations by the Magellan spacecraft confirm its validity remarkably well.

**Table 6:** Comparison of virial-action data with the Magellan spacecraft data for Venus

| Altitude km | $\delta T/\delta h$ x$10^5$ K/km | Virial temperature data K | Magellan temperature data[1] K | Virial pressure data x$10^6$ Pascals | Magellan pressure data[1] x$10^6$ Pascals | Virial density data x$10^{-20}$ cm$^{-3}$ | Magellan Density data[1] $p/kT = 1/a^3$ x$10^{-20}$ |
|---|---|---|---|---|---|---|---|
| 0  | 7.462 | 735.0 | 735 | 8.9918 | 9.330 | 8.8161 | 9.1938 |
| 5  | 7.706 | 697.7 | 697 | 5.7274 | 6.752 | 5.9459 | 7.0162 |
| 10 | 7.746 | 659.2 | 658 | 4.1810 | 4.800 | 4.5942 | 5.2834 |
| 15 | 7.788 | 620.4 | 621 | 2.9895 | 3.347 | 3.4900 | 3.9036 |
| 20 | 7.834 | 581.5 | 579 | 2.0868 | 2.281 | 2.5994 | 2.8553 |
| 25 | 7.884 | 542.3 | 537 | 1.4161 | 1.647 | 1.8913 | 2.2214 |
| 30 | 7.940 | 502.9 | 495 | 0.9289 | 1.087 | 1.3379 | 1.5905 |
| 35 | 8.003 | 463.2 | 453 | 0.5844 | 0.599 | 0.9139 | 0.9577 |
| 40 | 8.075 | 423.2 | 416 | 0.3489 | 0.355 | 0.5972 | 0.6181 |
| 45 | 8.160 | 382.8 | 383 | 0.1946 | 0.200 | 0.3682 | 0.3782 |
| 50 | 8.264 | 342.0 | 348 | 0.0991 | 0.108 | 0.2098 | 0.2248 |
| 55 | 8.390 | 300.7 | 300 | 0.0445 | 0.054 | 0.1071 | 0.1304 |
| 60 | 8.547 | 258.7 | 263 | 0.0166 | 0.024 | 0.0465 | 0.0661 |

[1]Jenkins et al., 1994

In all these computations, equation (25) is remarkably sensitive to any errors, omissions or choice of signs for different forms of energy. If the rotational symmetry of carbon dioxide is ignored and assigned a value of unity rather than the correct value of 2, or signs are confused, a highly erroneous plot of pressure and number density is obtained. Furthermore, application of equation (25) required that the chemical terms for vibration be additive, despite it being an internal energy. It has become clearer that this must be required by the singular nature of the virial for harmonic oscillators, where the mean kinetic energy and the mean potential energy are of equal value, rather than fractional as for translation and rotation. These choices of sign and magnitude were established empirically by trial and error, based on the resulting profiles. If incorrectly chosen, atmospheric pressures or number densities may even become inverted, increasing with altitude – which is impossible.

The relationship in equation (25) for carbon dioxide atmospheres has the consequence that the temperature lapse rate with altitude is variable, given that the factor $x$ is temperature dependent being equal to $h\nu/kT$, even at the relatively cool surface temperatures on Mars. For example, in Table 7 for carbon dioxide profiles on Mars, a lapse in temperature of 113.5 K for 30 km altitude is obtained with a surface temperature of 288 K, while the lapse for a surface temperature of 188 K is 116.9 K at 30 km altitude. Clearly, the vibrational entropy correction is required.

**Table 7:** Calculated thermodynamic profiles on Mars for a 95.32% $CO_2$ atmosphere

| Altitude (km) | Temp estimate K | Pressure estimate x$10^{-3}$ Pascals | Density estimate x$10^{-17}$ per cm$^3$ | Translational Gibbs energy as solved $kT\ln(n_t)^3 Q_e/z_t$ x$10^{13}$ergs per molecule | Rotational Gibbs energy estimate from $T$ $kT\ln[(n_r)^2/\sigma_r]$ x$10^{13}$ergs per molecule | Vibrational entropic energy x$10^{13}$ergs per molecule | Vibrational thermal energy from $T$ x$10^{13}$ergs per Molecule |
|---|---|---|---|---|---|---|---|
| | Warm | | | | | | |
| 0 | 288.0 | 11.0132 | 2.76836 | 8.222611 | 2.2472817 | 0.130172 | 0.188659 |
| 2.5 | 279.1 | 6.89733 | 1.78996 | 8.118458 | 2.1657591 | 0.056536 | 0.174673 |
| 5.0 | 270.1 | 5.49446 | 1.47322 | 7.911993 | 2.0839998 | 0.101783 | 0.143551 |
| 7.5 | 260.9 | 3.79194 | 1.05286 | 7.742523 | 1.9998844 | 0.088552 | 0.130079 |
| 10.0 | 251.7 | 2.57282 | 0.74050 | 7.572845 | 1.9168250 | 0.076431 | 0.117083 |
| 12.5 | 242.4 | 1.69588 | 0.50682 | 7.401234 | 1.8334629 | 0.065217 | 0.104396 |
| 15.0 | 233.0 | 1.08163 | 0.33627 | 7.227554 | 1.7497665 | 0.054922 | 0.092075 |
| 17.5 | 223.5 | 0.66443 | 0.21533 | 7.051694 | 1.6657228 | 0.045560 | 0.080186 |
| 20.0 | 213.9 | 0.39097 | 0.13282 | 6.873548 | 1.5813513 | 0.037141 | 0.068810 |
| 22.5 | 204.2 | 0.21895 | 0.07765 | 6.693029 | 1.4966575 | 0.029671 | 0.058036 |
| 25.0 | 194.4 | 0.11579 | 0.04314 | 6.510080 | 1.4116901 | 0.023149 | 0.047962 |
| 27.5 | 184.5 | 0.05729 | 0.02249 | 6.324692 | 1.3265235 | 0.175668 | 0.038692 |
| 30.0 | 174.5 | 0.02622 | 0.01088 | 6.136926 | 1.2412709 | 0.012900 | 0.033178 |
| | Cold | | | | | | |
| 0 | 188.0 | 11.0075 | 4.24088 | 5.090769 | 1.3562697 | 0.019410 | 0.042257 |
| 2.5 | 178.6 | 7.12265 | 2.88915 | 4.910867 | 1.2755107 | 0.014666 | 0.033303 |
| 5.0 | 169.0 | 4.41959 | 1.89386 | 4.727931 | 1.1945798 | 0.010719 | 0.025916 |
| 7.5 | 159.4 | 2.60511 | 1.18369 | 4.542952 | 1.1369869 | 0.007533 | 0.019469 |
| 10.0 | 149.7 | 1.44411 | 0.69868 | 4.356028 | 1.0329567 | 0.005637 | 0.014013 |
| 12.5 | 139.9 | 0.74367 | 0.38492 | 4.167437 | 0.9525243 | 0.003197 | 0.009572 |
| 15.0 | 130.1 | 0.35028 | 0.19498 | 3.977572 | 0.8726106 | 0.001887 | 0.006130 |
| 17.5 | 120.3 | 0.14783 | 0.08903 | 3.786927 | 0.7934571 | 0.001021 | 0.003623 |
| 20.0 | 110.4 | 0.05438 | 0.03568 | 3.596082 | 0.7153219 | 0.000496 | 0.001934 |
| 22.5 | 100.5 | 0.01679 | 0.01210 | 3.405643 | 0.6384594 | 0.000209 | 0.000906 |
| 25.0 | 90.7 | 0.00412 | 0.00329 | 3.216181 | 0.5630984 | 0.000074 | 0.000357 |
| 27.5 | 80.9 | 0.00074 | 0.00066 | 3.028166 | 0.4894292 | 0.000002 | 0.000111 |
| 30.0 | 71.1 | 0.00008 | 0.00008 | 2.841940 | 0.4176100 | 0.000000 | 0.000025 |

Profiles were plotted at arbitrary temperatures using pressure and density data from NASA (2014).

**Discussion**

*The adiabatic lapse rate* **versus** *the virial-action lapse rate*

Currently, climatologists place reliance on the concept of adiabatic processes in which no heat exchange occurs by a parcel of air subject to convection. Adiabatic processes imply there is no change in entropy. Temperature changes for adiabatic processes such as convection can then be estimated (Rogers, 1976) appealing to the ideal gas law and the potential temperature $\Theta$. This is the temperature that a parcel of air at its current state would have if it were subjected to an adiabatic compression or expansion to a final pressure of $10^5$ pascals or 1 bar, close to 1 atmosphere (1.013 bar).

$$\Theta = T(100000/p)^k$$

Here $k$ is equal to a function of constant pressure and constant volume heat capacities being $(C_p - C_v)/C_p$ or $R/C_p$, equal to 0.286 for dry air at ambient temperatures. Thus, for air at 5 km altitude where the virial theorem indicates the temperature would be equal to 253.7 K (see Table 3) and pressure is $0.5286 \times 10^5$ pascals, $\Theta$ can be calculated using the formula above as equal to 304.5 K. However, this adiabatic compression would generate a thermodynamic surface pressure of 1.057 atmospheres, failing to agree with the gravitational pressure of 1 atmosphere. It also indicates an adiabatic lapse rate of around 10.16 K per km. This high rate compares to a virial lapse rate calculated for air in Table 3 of 6.89 K per km. It is clear from the USAF data in Table 3 that the actual lapse rate observed of 6.50 K per km up to 5 km is much closer to the virial lapse rate than an adiabatic rate. Humid air is expected to have a lower lapse rate because of the release of latent heat of vaporization as the temperature falls.

While pseudo-adiabatic processes in the atmosphere where entropy is largely conserved occur, because rapid convective processes do not allow heat to equilibrate, the performance of gravitational work consuming heat must modify their nature. In any case, an adiabatic expansion only cools a gas if external work is being done as expansion into a vacuum has no capacity for cooling *per se*. Conversely, an adiabatic compression heating air must be thought of as the performance of work by gravity on the parcel of air. The heating of air as it descends in the high pressure zone of an anticyclone is indicative of a compressive process, resembling the heating of air in a bicycle pump as work is performed. In convecting air it is impossible for all the molecules in a parcel to ascend without a simultaneous diminution of pressure accompanied by lateral transfer of air. This implies that there must also be a reduction of air at all heights in the ascending column, implying spiral flows and turbulence.

Considering an isoentropic process using action theory, we have the following equation that must not vary for a molecule like nitrogen.

$$s = k\ln[e^{7/2}(@_t/\hbar)^3(@_r/\hbar)^2 Q_e/(z_t\sigma_r)]$$

Thus an isoentropic process in air requires the following equality.

$$k\ln[e^{7/2}(@_{to}/\hbar)^3(@_{ro}/\hbar)^2 Q_e/(z_t\sigma_r)] - k\ln[e^{7/2}(@_{tn}/\hbar)^3(@_{rn}/\hbar)^2 Q_e/(z_t\sigma_r)] = 0$$

$$k\ln[(@_{to}/\hbar)^3(@_{ro}/\hbar)^2] - k\ln[(@_{tn}/\hbar)^3(@_{rn}/\hbar)^2] = 0$$

Thus $(@_{to}/\hbar)^3(@_{ro}/\hbar)^2 = (@_{tn}/\hbar)^3(@_{rn}/\hbar)^2$

$$@_{to}^3 @_{ro}^2 = @_{tn}^3 @_{rn}^2$$

Clearly, if the translational action increases rotational action must decrease to the power 0.67 for the entropy to remain the same. Thus, even if we allow the entropy to vary as a result of gravitational work being performed, the change in rotational free energy is subtracted from the change in translational free energy, consistent with equations (21) and (22).

We can also write, using the formulae used to calculate translational and rotational action,

$$(3kT_o I_{to})^{3/2}(2kT_o I_{ro}) = (3kT_n I_{tn})^{3/2}(2kT_n I_{rn})$$

$$(T_o)^{5/2}(I_{to})^{3/2} I_{ro} = (T_n)^{5/2}(I_{tn})^{3/2}(I_{rn})$$

$$(I_{to}/I_{tn})^{3/2} = I_{rn}/I_{ro}(T_n/T_o)^{5/2} = (r_o/r_n)^3$$

Given that the rotational moment of inertia does not change appreciably with temperature, the ratio of the translational moments of inertia ($I_{to}/I_{tn}$) must decrease by the change in temperature to the power 1.67. That is the following relationship would hold for an isoentropic or adiabatic process.

$$I_{to} T_o^{5/3} = I_{tn} T_n^{5/3}$$

However, despite the fact that the translational entropy increases with height while the rotational entropy decreases with height, for the data in Table 3 for the Earth's atmosphere, the former increase is about four times the magnitude of the latter. Overall, there is a steady increase in the entropy with the height of a steady-state equilibrium atmosphere. Despite the fact that the adiabatic model does provide an approximate guide to atmospheric physics and the causes of weather, it is suggested that use of the virial-action model as a basis for calculating temperature gradients and accounting for their convective modification upwards above 6.9 K per km or by water condensation in cloud formation downwards from 6.9 K per km on would be more accurate. Climate scientists are invited to test this hypothesis.

*The atmosphere as a recycling heat engine*

There is a subtle yet decisive difference between an adiabatic expansion of the working substance in the cylinder of a heat engine as proposed by Carnot and that of a parcel of air subject to convection in the atmosphere. In the Carnot engine the working fluid is maintained at almost the same gravitational potential during the expansion, so all internal changes are considered as thermal. The fluid cools dependent on external work being performed during the isoentropic phase requiring heat. This cooling is reversed during the subsequent isoentropic compression restoring the fluid to the original state. By contrast, in the atmospheric convection cycle, the working fluid itself is the object of the work performed by moving to a higher gravitational potential, accompanied by expansion to a lower pressure, thus conforming with the ideal gas law (equation 17). In the Carnot cycle, the total number of molecules in the working fluid remains constant, with a uniform reduction in number density proportional to the expanded volume during adiabatic expansion. In the atmospheric cycle, the expansion of the parcel of air must be matched by a simultaneous increase in the number density of an adjoining descending parcel of air, allowing space for the convective expansion. The expansion is contingent on a compression, inferring that some gas molecules are transferred horizontally at all altitudes to allow the correct number density and pressure to be maintained.

The model presented here aims to provide a reliable causal basis for modeling the dynamic motions of the atmosphere. It proposes that the driving forces for convection and advection are thermodynamic effects, a result of molecular forces tending to equilibrium with gravitation. However, in the global weather system, it is clear that this only approaches but never reaches true equilibrium since it depends on heat flow. For equilibrium with height there must be exact compensation between the increase in gravitational molecular potential *mgh* and the decrease in molecular action potential, as shown by the increase in action and entropy; because of their greater heat capacity and internal energy, including vibration, greenhouse molecules are even more capable of decreasing action and increased action potential with height, as they become colder. Even for nitrogen and oxygen, the main constituents of the atmosphere, these opposing variations in translational and rotational molecular energy contribute strongly to the increased gravitational potential achievable with height – this internal/external action reversal with altitude being the efficient cause of the cooling effect with height.

In this virial-action model, the cooling of the atmosphere with altitude is not primarily caused by radiation to space as is sometimes assumed. This would confuse cause with effect. On the contrary, it is an obligatory thermodynamic result characteristic a heat engine doing work, allowing the atmosphere to reach its correct altitude where it can radiate at the correct temperature, consistent with the Stefan-Boltzmann equation. This apparent heating at the surface above the Earth's black body temperature of about 254 K, or failure to cool, is not a greenhouse effect as commonly understood – but a trapping of heat required to equilibrate the thermodynamic pressure as given by the ideal gas law with the pressure of air from its weight as determined by gravity (see equation (17)). This conclusion provides an alternative viewpoint to the effect of greenhouse gases on global warming.

Because thermal equilibrium is reached relatively slowly, much of the atmosphere's weather occurs under non-equilibrium conditions. Thus, the cellular nature of the atmosphere's motion of cyclones and anticyclones and descending high pressure and ascending lows pressure zones is a result of thermodynamic gradients, primarily caused by gradients in pressure and temperature. Using the virial-action model for calculating entropy and free energy, it will be relatively easy to estimate the steepness of these gradients and propensity for climatic action and weather.

*Coupling of molecular thermodynamics to radiative transfers in the atmosphere*
Climate models suggest that radiative transfers in the infrared zone of frequencies can be considered as subject solely to the Stefan-Boltzmann law for black body radiation, independent of gravity. However, according to the theory presented here, radiation must be directly coupled to molecular thermal motions affected primarily by gravity, satisfying the thermodynamic gradients identified for the different gases in this paper (see Tables 2-6). Not only does the kinetic energy decline, as shown by the decreased temperature, but so does the capacity of the parcel of air to emit radiation, subject to the Stefan-Boltzmann law. As temperature falls from heat performing gravitational work, longer wavelengths directly proportional to the lower temperature will be emitted. The intensity of radiation at each altitude must depend strictly on the temperature and is suggested to be independent of the gaseous composition.

Emissions of infrared quanta will be mediated by molecular collisions and the second law tendency of thermodynamics to satisfy equilibrium. Furthermore, ascending masses of air will absorb heat radiation as hydrostatic pressure is reduced, effectively converting thermal energy into translational and then gravitational energy, the air now radiating energy of longer wavelengths. By contrast, descending masses of air in zones of increased pressure will progressively radiate shorter, hotter wavelengths, as gravitational energy is released as thermal energy.

So from action theory, the atmospheric gases can be considered as having their elevation continuously supported by transfers of thermal energy, countering gravity. The greater weight and pressure at the surface demands a higher temperature. In descending air, the molecular kinetic energy increases as gravitational potential declines, also a virial effect. This warming does not require solar heat as the Earth's surface compresses the air (the "bicycle pump effect"), generating kinetic energy gravitationally at the same time as the total field energy increases, shown as increased thermal radiation. This gravity induced thermodynamic heating effect of dehydrated descending air temporarily independent of the sun is an important antecedent for wild-fires.

If the Sun's thermonuclear fusion was extinguished and solar radiation ceased, the Earth's atmosphere would gradually lose its heat to space and cool, first condensing as a liquid below $100^o$ K and then ultimately freezing in a layer on the surface less than 10 metres thick, gravity taking full control of the atmosphere's morphology. On solar re-ignition, the reverse process would occur, with radiant energy from the Sun absorbed at the Earth's surface producing heat then being converted into colder gravitational energy, cooling the surface atmosphere as thermal energy is reabsorbed, increasing gravitational potential energy as the atmosphere rises. The equations given in this paper are exactly consistent with the magnitude of these two effects, the declining kinetic energy and the absorbed radiation being equal. The increased gravitational potential energy is the sum of these two effects, as the virial theorem predicts.

The equations given before indicate the steepness of the thermodynamic gradient. For example, equation (9) for monatomic gases can easily be expressed as a counter force.

$$mg = \{kT_o\ln[(\hbar/@_{to})^3 z_t/e] - kT_n\ln[(\hbar/@_{tn}/\hbar)^3 z_t/e] - 3/2k(T_o - T_n)\}/h_n$$

Calculations based on such thermodynamic forces for parcels of air differing in potential because of temperature or pressure differences can easily be made using the action thermodynamics model. Its essence is the counterpoising of thermodynamics and gravity. This is most clearly evident in the chaotic variations of daily weather.

As a result of the absorption of radiation from the Earth's surface within the atmosphere, the internal vibrational and rotational actions and entropies of atmospheric greenhouse molecules will increase. The inertial forces generated will then directly transfer energy to translational action in collisions, increasing the translational entropy and lowering the local pressure by raising the atmosphere. Paradoxically, the greenhouse molecules with greater internal action cause the atmosphere to rise to a greater height and to become cooler – a direct response to their internal entropy at base levels and their greater capacity to absorb the Earth's radiation as temperature falls, giving a lower lapse rate of temperature with height. An atmosphere of argon (see Table 2) would be even less elevated than an atmosphere of carbon dioxide, despite the greater mass of the latter, since argon lacks the enthalpic effect of internal rotational and vibrational energy that reduces the virial lapse rate driving the atmosphere higher as indicated by equation (22). However, being of similar total mass, both such theoretical atmospheres would cool to around 254 K, the Earth's temperature from space, but at only three kilometres rather than almost six (see Tables 2 and 3). This hypothesis can readily be tested, both in theory and practice.

The capacity of the atmosphere to absorb thermal energy from the Earth's surface by lifting the atmosphere, thus expanding its volume, is ultimately only limited by the tendency of its gases to be dissociated from Earth's gravitational attraction. Thus, the atmosphere is far from being a uniform static body with a defined heat capacity per unit volume (Kennedy 2001, Ch. 5); as a heavier gas, carbon dioxide should be bound to the Earth more tenaciously, but its internal rotational and vibrational energy improves its capacity to lift the atmosphere. Its greenhouse properties should also contribute to greater inertia of motion of advective air cells and thus the power of winds. Understanding these features of greenhouse gases is essential for effective management of their environmental impacts.

Through feedback from the equilibration of pressures, the surface temperature will depend on rates of solar absorption at the surface and the statistical capacity to emit radiation to space as affected by the presence of gases in the atmosphere. Ultimately, the Earth will emit radiation with an intensity adequate to match the insolation absorbed from the Sun, with the atmosphere raised to a height that ensures that, statistically, this emission occurs at the correct black body temperature of $254^o$ K. This temperature occurs near 5 km on Earth as shown in Table 3. Overall, this is a self-organising physical system partly reminiscent of Lovelock's Gaia (1979). However, it depends on the physical mechanism of equal gravitational and thermodynamic pressure (equations (17) and (22)) with the ideal gas law determining the equilibrium temperature. This strong tendency of equality in gravitational and thermodynamic pressures must apply at all altitudes, with corresponding adjustments in number density of gas molecules and temperature.

At a given rate of insolation and albedo, a major determinant identified here of the average surface temperature and its difference from the black body temperature as estimated by the Stefan-Boltzmann law will be the weight of the atmosphere. This is well illustrated by the moderate surface warming of 33-34 K on Earth, an extreme 560 K on Venus and the negligible amount on Mars, despite Mars having an 8-fold higher carbon dioxide pressure than Earth (NASA, 2014). Although Venus is often claimed to have a "runaway greenhouse effect", the warming is no more than expected, given the greater weight of its atmosphere and the need to have a higher temperature and thermal energy content as indicated by the Stefan-Boltzmann Law to support its weight.

In that sense, the Earth's mean surface temperature 33-34$^o$ K warmer at 288$^o$ K expresses an internal temperature gradient sufficient to continuously do necessary thermodynamic work on elevating the atmosphere. This is required to allow it to emit radiation statistically averaged to 254$^o$ K. Obviously, to achieve this mean value, the surface must be warmer. The same formulae derived in this paper can be used at all latitudes and surface temperatures and pressures simply by varying the inputs to the computer program for surface pressure and temperature. This virial-action hypothesis provides a thermodynamic basis for the greenhouse effect. However, the gravitational pressure is not a function of the atmosphere's composition

– only its weight. By contrast, the thermodynamic pressure response does vary with the atmosphere's composition, as this paper shows in the profiles constructed in Tables 2-6 with different gas compositions. These pressure relationships require further study to understand their effect on climate change.

The surface temperature with full emissivity indicates that the Earth's surface radiates more heat per unit area from its surface (ca. 390 Joules sec$^{-1}$ m$^{-2}$) than it receives and absorbs from the Sun (235 Joules sec$^{-1}$ m$^{-2}$). By contrast, the Earth emits to space about 70% of its previously absorbed shorter wave solar radiation from nearer the top of the elevated atmosphere, although the average wavelength is more than 20 times as long, near 12 μm in the infrared. Only some 30% of the longer wave radiation is regarded as emitted directly from the surface to space without doing work on the atmosphere. Of course this process of elevating the atmosphere to keep it cooler than the surface is a dynamic approach to equilibrium that is never achieved, because of the varying rates of insolation by day and night, season and latitude and solar activity.

*Assessing climatic risks from human activities*

This paper does not dispute general conclusions regarding anthropogenic effects on global warming. But how good is the evidence that the products of fossil fuels and their radiative effects are the sole factors causing the change? Might it not be that the thermodynamic mechanisms identified in this paper are now lifting the atmosphere higher, effectively raising its heat capacity and thus cooling it to the extent required for black body radiation at the tropopause? Certainly, it is possible the temperature of the ocean will be responsive to changes in local atmospheric pressure as a result of compressive work being done in air and on the ocean, releasing latent heat – the 'bicycle-pump effect'. Has there have been too much reliance on radiative effects as the cause of global warming, given that compression of circulating air masses during surface friction releasing heat from the statistical component of atmospheric entropy could also provide this result, from many human activities?

The success of the thermodynamic approach used here to calculate atmospheric profiles suggests that radiative balance between Earth's absorption of shortwave solar energy and its subsequent emission as long wave radiation to space should be spontaneously achieved in the tendency to hydrostatic equilibrium. Indeed, the virial-action model claims that the atmosphere will spontaneously adjust its altitude thermodynamically, using heat to drive cooling gravitational work (and *vice versa*) as shown in the tables, achieving the correct balance of colder radiation to space. The need to equate the thermodynamic pressure with the gravitational pressure generates surface warming just balancing the pressure exerted by the weight of the atmosphere, even for non-greenhouse gases.

**Conclusion**

Action thermodynamics shows how the atmosphere stays cooler by convectively increasing its gravitational potential whenever its thermal energy content near the surface increases. A direct relationship between the surface temperature, the lapse rate and the height of the atmosphere can be expected as a function of these thermodynamic forces. Assuming heat transfer can proceed swiftly enough from the surface to achieve a steady-state thermal equilibrium, any increase in surface temperature is automatically counterbalanced by raising and cooling the atmosphere – and *vice versa*. That this virial-action model correctly predicts the atmospheric profiles actually observed on Earth, Venus and Mars as shown in the tables strongly suggests that a steady-state thermal equilibrium with gravity is obtained, subject to the range of variation constituting weather.

However, there is a need to further analyse the effect of substituting virial-action to replace adiabatic lapse rates as the primary cause of temperature gradients with altitude. The new theory presented here does not dispute that convection and phase changes such as condensation of water or carbon dioxide will modify temperature gradients. But real temperature changes such as in the descending high pressure zones will be more accessible if the correct basic physics is employed as background. If the virial-action hypothesis is valid most of the temperature change with altitude requires no convection of parcels of air or effects of condensation. The success in this paper in demonstrating self-organisation of planetary atmospheres in thermodynamic terms – even involving obvious quantum effects, particularly for carbon dioxide on Venus – should be compelling.

The virial-action hypothesis can be further tested on Jupiter and Saturn at colder temperatures with mainly hydrogen in their atmospheres. A role for greenhouse gases in determining the dynamics of the atmospheric profile is included in the model. Such greenhouse gases will change the lapse rate in temperature with height, although this effect is offset by their greater mass. This provides us an alternative model of the atmosphere that can be applied to help manage the impacts of climate change.

This action revision is based on recognised thermodynamic principles, scientific fields largely initiated by Clausius (1850) and Gibbs (1902), from Carnot's powerful beginning discussing heat engines. It could therefore give more reliable predictions regarding global warming and climate change than the more descriptive general circulation models currently employed (Harvey, 2000; Manabe, 1998). It is recommended that new climate models be developed that link gravity to thermodynamics, radiation to space and climate to allow critical experimental tests with the exclusive rigor recommended by the late Karl Popper – a key inspiration for this work. We may never reach the whole truth but at least we can eliminate our errors, above all avoiding attempts to confirm them (Popper, 1972), as we progress on this journey of discovery.

**Acknowledgements**

I am grateful for support from colleagues, Rodney Roughley, Michael Rose, Angus Crossan, Michael Roderick, Barry Noller, John Knight and many others. This work was conducted without specific financial support.

**References**

Anderson, G.P., Clough, S.A., Kneizys, F.X., Chetwynd, J.H. and Shettle, E.P. (1986) *AFGL Atmospheric Constituent Profiles*, (0-120 km), Report of Air Force Geophysics Laboratory, Project 7670, Hanscom, USA.

Berberan-Santos, M.N., Bodunov, E.N. and Pogliani, L. (1997) On the barometric formula. *American Journal Physics* 65, 404-412.

Boltzmann, L. (1896) in J.A. Barth, *Lectures in Gas Theory.* Translated by Stephen G. Brush, p. 342, 1964, Dover Publications, New York.

Brown, Arthur (1968) *Statistical Physics.* The University Press, Edinburgh.

Clausius, R. (1850) On the motive power of heat, and on the laws which can be deduced from it for the theory of heat. Reproduced in the Dover Books E. Mendoza (1988) edition of Sadi Carnot (1824) *Reflections on the motive power of fire* from *Poggendorff's Annalen der Physik* LXXIX 368, 500.

Clausius, R.J.E. (1870) On a mechanical theorem applicable to heat. *Philosophical Magazine, Series* 4, 40, 122–127.

Clausius, R. (1875) *The Mechanical Theory of Heat.* Translated by Walter R. Browne, 1879 MacMillan and Co., London.

Gibbs, Josiah Willard (1902) *Elementary Principles in Statistical Mechanics.* Charles Scribner's Sons, New York.

Glasstone, S. (1951) *Textbook of Physical Chemistry*, Macmillan and Company, London.

Harvey, L.D.D. (2000) *Global Warming The Hard Science*, Prentice Hall Harlow UK.

Jain, A.K., Briegleb, B.P., Minschwaner, K. and Wuebbles, D.J. (2000) Radiative forcings and global warming potentials of 39 greenhouse gases. *Journal of Geophysical Research* 105, 20,773-790.

Jenkins, J.M. , Steffes, P. G. , Hinson, D.P. , Twicken, J.D. and Tyler, G.L. (1994) Radio occultation studies of the Venus atmosphere with the Magellan spacecraft, *Icarus* 110, 79-94.

Kennedy, I.R. (2000) Action as a dynamic property of the *genotype x environment* interaction: Implications for biotechnology. *Acta Biotechnologica* 20, 351-368.

Kennedy, I.R. *Action in Ecosystems: Biothermodynamics for Sustainability.* Research Studies Press, 251 pages, Baldock, UK, 2001.

Kennedy, Ivan R., Harold Geering, Michael T. Rose and Angus N. Crossan (2006-2013) Modelling the entropy and free energy of atmospheric greenhouse gases from their action. Draft available on request.

Lovelock, J.E., (1979) *The Ages of Gaia: A Biography of our Living Earth.* Norton, New York USA.

Manabe, S. (1998) Study of global warming by GFDL climate models. *Ambio* 27,182-86.

Mitchell, J.F.B. (1989) The "Greenhouse" effect and climate change. *Reviews of Geophysics* 27,115-139.

Moore, W. (1962) *Physical Chemistry*, Longmans, London.

Morowitz, H.J. (1978) *Foundations of Bioenergetics* (see pages 142-145). Academic Press, New York.

Perrin, M. Jean (1909) *Brownian Movement and Molecular Reality.* Soddy, Taylor and Francis, London 1910.

Popper, K.R. (1972) *Objective Knowledge: An Evolutionary Approach.* Oxford University Press, London.

Rogers, R.R. (1976) *A short course in cloud physics,* p. 8, Pergamon Press Oxford.

Rose, M.T., Crossan, A.N. and Kennedy, I.R. (2008) Sustaining action and optimising entropy: coupling efficiency for energy and the sustainability of global ecosystems. *Bulletin of Science, Technology and Society*, 28, 260-272.

Venus Fact Sheet, National Aeronautics and Space Agency, accessed March 1, 2014, http://nssdc.nasa.gov./planetary/factsheet.html; Mars Fact Sheet, National Aeronautics and Space Agency, accessed March 1, 2014 http://nssdc.nasa.gov./planetary/factsheet.html

UNEP (2013) *Climate Change 2013. The Physical Science*. Report of the Intergovernmental Panel on Climate Change, Working Group 1 – 12th Session, Stockholm, 23-26 September 2013.

Weast, R.C. (1969) *CRC Handbook of Chemistry and Physics*, Table 1-10, 50th edition, Chemical Rubber Company, Boca Alton Florida.